\documentclass[notitlepage,floatfix,aps,prx,reprint,amsmath,amssymb,twocolumn,superscriptaddress,10pt]{revtex4-2}
\usepackage{graphicx}
\usepackage[pretty,uselistings]{revquantum}
\usepackage{soul}

\usepackage{xcolor}

\begin{document}
\title{Coherent Control of Quantum-Dot Spins with Cyclic Optical Transitions}
% \author{Authors}
\author{Zhe Xian Koong}
\email[Corresponding author: ]{zxk22@cam.ac.uk}
\affiliation{
 Cavendish Laboratory, University of Cambridge, JJ Thomson Ave, Cambridge CB3 0HE, United Kingdom
}
\author{Urs Haeusler}
\affiliation{
 Cavendish Laboratory, University of Cambridge, JJ Thomson Ave, Cambridge CB3 0HE, United Kingdom
}
\author{Jan M. Kaspari}
\affiliation{
 Department of Physics, TU Dortmund University, 44221 Dortmund, Germany
}
\author{Christian Schimpf}
\affiliation{
 Cavendish Laboratory, University of Cambridge, JJ Thomson Ave, Cambridge CB3 0HE, United Kingdom
}
\author{Benyam Dejen}
\affiliation{
Cavendish Laboratory, University of Cambridge, JJ Thomson Ave, Cambridge CB3 0HE, United Kingdom
}
\author{Ahmed M. Hassanen}
\affiliation{
 Cavendish Laboratory, University of Cambridge, JJ Thomson Ave, Cambridge CB3 0HE, United Kingdom
}
\affiliation{
Department of Engineering Science, University of Oxford, Parks Road, OX1 3PJ, United Kingdom
}
\author{Daniel Graham}
\affiliation{
 Cavendish Laboratory, University of Cambridge, JJ Thomson Ave, Cambridge CB3 0HE, United Kingdom
}
\author{Yusuf Karli}
\affiliation{
 Cavendish Laboratory, University of Cambridge, JJ Thomson Ave, Cambridge CB3 0HE, United Kingdom
}
\author{Ailton J. Garcia Jr.}
\affiliation{Institute of Semiconductor and Solid State Physics, Johannes Kepler University, Linz,
Austria}
\author{Melina Peter}
\affiliation{Institute of Semiconductor and Solid State Physics, Johannes Kepler University, Linz,
Austria}
\author{Edmund Clarke}
\affiliation{EPSRC National Epitaxy Facility, University of Sheffield, Broad Lane, Sheffield S3 7HQ, United Kingdom}
\author{Maxime Hugues}
\affiliation{Université Côte d’Azur, CNRS, CRHEA, rue Bernard Gregory, 06560 Valbonne, France}
\author{Michał Gawełczyk}
\affiliation{Institute of Theoretical Physics, Wrocław University of Science and Technology, 50-370 Wrocław, Poland}
\author{Armando Rastelli}
\affiliation{Institute of Semiconductor and Solid State Physics, Johannes Kepler University, Linz,
Austria}
\author{Doris E. Reiter}
\affiliation{
 Department of Physics, TU Dortmund University, 44221 Dortmund, Germany
}
\author{Mete Atat{\"u}re}
\email[Corresponding author: ]{ma424@cam.ac.uk}
\affiliation{
 Cavendish Laboratory, University of Cambridge, JJ Thomson Ave, Cambridge CB3 0HE, United Kingdom
}
\author{Dorian A. Gangloff}
\email[Corresponding author: ]{dag50@cam.ac.uk}
\affiliation{
 Cavendish Laboratory, University of Cambridge, JJ Thomson Ave, Cambridge CB3 0HE, United Kingdom
}
\date{\today}

\begin{abstract}
Solid-state spins are promising as interfaces from stationary qubits to single photons for quantum communication technologies. 
Semiconductor quantum dots have excellent optical coherence, exhibit near-unity collection efficiencies when coupled to photonic structures, and possess long-lived spins for quantum memory. 
However, the incompatibility of performing optical spin control and single-shot readout simultaneously has been a challenge faced by almost all solid-state emitters.
To overcome this, we leverage light-hole mixing to realize a highly asymmetric lambda system in a negatively charged heavy-hole exciton in Faraday configuration. By compensating GHz-scale differential Stark shifts, induced by unequal coupling to Raman control fields, and by performing nuclear-spin cooling, we achieve quantum control of an electron-spin qubit with a $\pi$-pulse contrast of $97.4\%$ while preserving spin-selective optical transitions with a cyclicity of $471\,(50)$. We demonstrate this scheme for both GaAs and InGaAs quantum dots, and show that it is compatible with the operation of a nuclear quantum memory. Our approach thus enables repeated emission of indistinguishable photons together with qubit control, as required for single-shot readout, photonic cluster-state generation, and quantum repeater technologies. 
\end{abstract}

\maketitle

\section{Introduction}
% Distributed quantum information processing relies on matter-based nodes with long memory time becoming entangled via optical photons. 
Quantum network nodes based on efficient, long-lived spin-photon interfaces are essential for distributed quantum information processing~\cite{kimble_the_2008,ritter_an_elementary_2012, ruf_quantum_2021}.
Each node consists of an interface between single photons and a spin qubit, and realizing its full functionality requires initializing and reading out the spin state, encoding information via quantum control over the spin, entangling the spin with a photon, and storing this quantum information in a local memory before sending the photon across the optical link~\cite{duan_long_2001,sangouard_quantum_2011}. These requirements typically reduce to the simultaneous ability to coherently control a long-lived spin ground state, and to trigger spin-selective emission of photons via cyclic transitions. While direct magnetic-dipole control of the ground state spin is possible in several systems with cyclic transitions~\cite{Gritsch_Optical_2025, Raha_Optical_2020,Stas2022a}, in many quantum emitter systems this remains an outstanding problem to integrate with efficient photonic structures~\cite{Guo2023}, and all-optical means are thus the only viable route for quantum control. However, all-optical control requires linking the spin states via multiple allowed optical transitions, which can conflict directly with the need for cyclic optical transitions.

Amongst quantum emitters, semiconductor quantum dots (QDs) stand out for their exceptional optical coherence and efficiency, thanks to the continued improvement in material quality~\cite{senellart_high-performance_2017, zhai_quantum_2022} and photonic technologies~\cite{tomm_abright_2021,Appel_Coherent_2021}. They also feature coherent electronic and nuclear spins~\cite{Gangloff_quantum_2019,Chekhovich2020,shofer_tuning_2024,Appel2025}, such that 100-microsecond~\cite{Zaporski_Ideal_2023} and 100-millisecond~\cite{Dyte2025} coherence times are attainable for GaAs QDs, respectively. 
While some low-field control of the spin is possible with limited fidelity~\cite{Huet2025}, addressing and controlling the electronic spin as a high-fidelity qubit typically requires a few-Tesla magnetic field to lift the degeneracy of the electronic spin states beyond the scale of hyperfine and exciton linewidth energies. 
Under a growth-axis field (Faraday configuration), the presence of spin-specific circularly-polarized cyclic transitions allows for efficient spin-state readout~\cite{nick_vamivakas_spin-resolved_2009,delteil_observation_2014} in the single-shot regime~\cite{Vamivakas2010,Antoniadis_cavity_2023}. 
However, spin control of a QD in this configuration has not been realized. 
Direct microwave drive would need to operate in the few to tens of GHz regime and would need to reach Rabi frequencies of tens of MHz at minimum  -- for nuclear-spin stabilization of the qubit splitting~\cite{jackson_optimal_2022} -- corresponding to tens of mT of magnetic flux density at the QD position; this is a formidable challenge for optically active QD systems. The historical and viable approach for spin control in a QD has thus been to apply an in-plane field (Voigt configuration), resulting in a symmetric linearly polarized $\Lambda$ system which allows all-optical coherent control over the spin states ~\cite{press_complete_2008,Gangloff_quantum_2019,Bodey_Opitcal_2019,Jackson_quantum_2021,jackson_optimal_2022,Hogg2025}, reaching high contrast ($99.3$\%~\cite{Zaporski_Ideal_2023}) and full-phase control of an electronic spin with GHz-scale Rabi frequency~\cite{Hogg2025}. Despite efforts to selectively enhance a readout transition in this configuration, via the coupling to a cavity~\cite{Sweeney_cavity_2014, Hogg2025} or a waveguide~\cite{Appel_Coherent_2021,Meng2024}, only modest cyclicity $\sim 36$ has been achieved. 
The lack of intrinsically cyclic transitions in the Voigt configuration has hampered the simultaneous realization of both spin-selective readout and all-optical spin-control, and has been a major hurdle for QD systems vying for the realization of photonic quantum technologies.

\begin{figure*}
    \centering
    \includegraphics[width=1\textwidth]{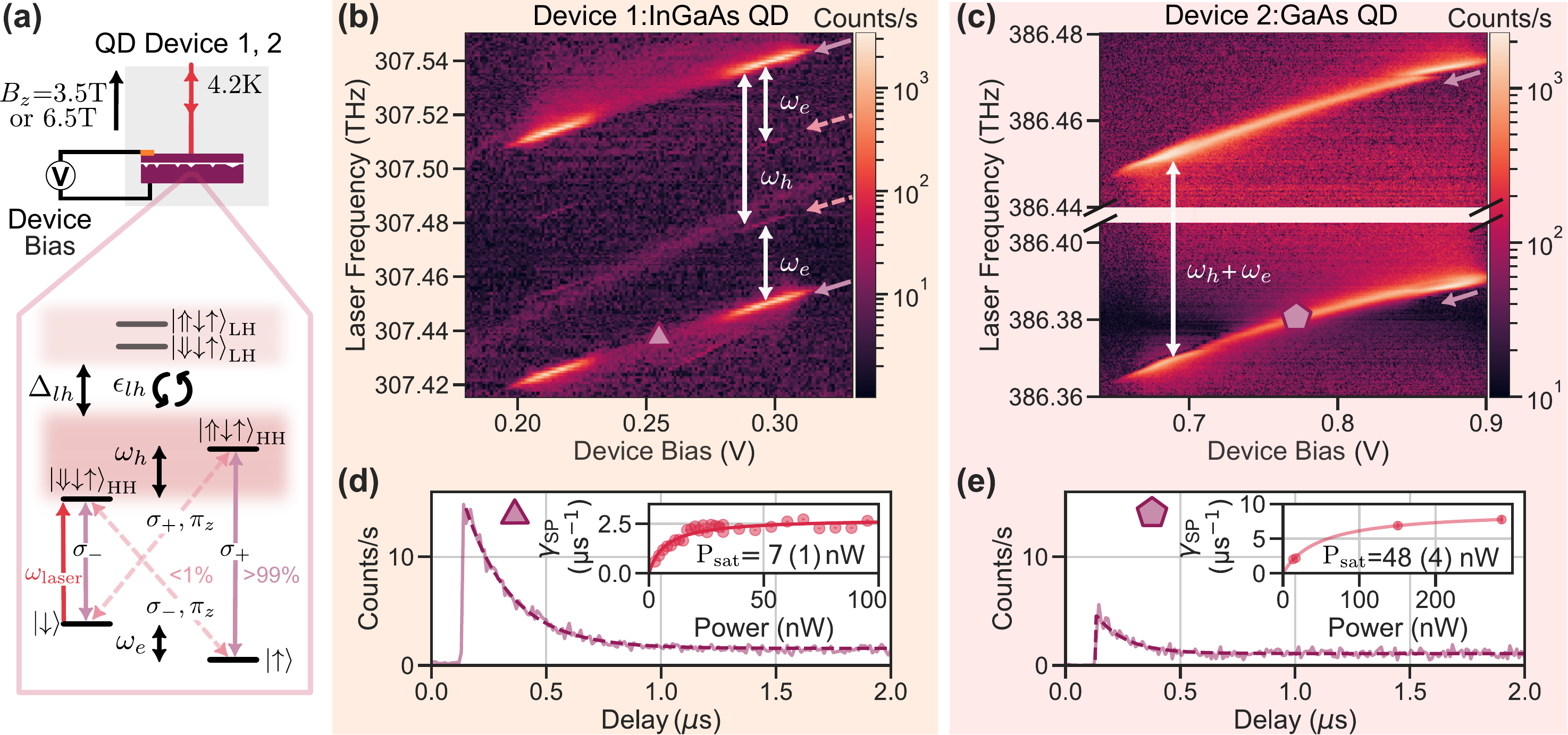}
    \caption{\textbf{Optical Cyclicity of InGaAs and GaAs QDs in Faraday Geometry.}
    \textbf{(a)} Top panel: schematic of experimental setup (c.f. Appendix~\ref{app:hardware}) consisting of a charge-tuneable device with bias voltage $V$, containing QDs, under a  magnetic field $B_z$ up to 6.5~T along the QD device growth axis $z$. 
    % A voltage bias $V$ is applied to the Schottky (Device 1) or to the p-i-n (Device 2) diode to enable charge control and tuning of the emission energy via the d.c. Stark shift. 
    Bottom panel: energy levels of a negatively-charged exciton $X^{1-}$ in Faraday configuration for a negative electron $g$-factor~\cite{Schimpf_Optical_2025}. 
    Solid purple arrows show the dominant circularly polarized dipole transitions connecting the electron-spin ground states, $\ket{\downarrow}$ and $\ket{\uparrow}$, to the HH trion states, $\ket{\Downarrow\downarrow\uparrow}_\mathrm{HH}$ and $\ket{\Uparrow\downarrow\uparrow}_\mathrm{HH}$, respectively. 
    A small admixture ($\epsilon_{lh}$) of the LH $X^{1-}$ states ( $\ket{\Downarrow\downarrow\uparrow}_\mathrm{LH}$ and $\ket{\Uparrow\downarrow\uparrow}_\mathrm{LH}$), split to higher energy by $\Delta_{lh}$, with the HH states enables the spin-flipping transitions (pink dashed arrows) to the two electron ground states.
    % The branching ratio of the spin-conserving (spin-flipping) transitions is typically $>99\%$ ($<1\%$).
    \textbf{(b)} (\textbf{(c)}) Resonant spectroscopy of the $X^{1-}$ of an InGaAs (GaAs) QD at $B_z=3.5\,$T with a linearly polarized narrow-linewidth laser at a power above saturation, $\sim 2\mathrm{P_{sat}}$ ($\sim 8\,\mathrm{P_{sat}}$), as a function of the device bias $V$ and laser frequency. 
    The spin-conserving transitions are highlighted by solid purple arrows, while the spin-flipping transitions are highlighted by pink dashed arrows.
    For the InGaAs QD (Device 1), we extract an electron splitting $\omega_e=30(1)\,$GHz and a hole splitting $\omega_h=59(1)\,$GHz from Lorentzian fits to the resonances at $V=0.3$~V.
    The combined electron and hole splitting is $\omega_e + \omega_h = 81.6(2)\,$GHz for the GaAs QD (Device 2).
    % The discontinuity in the GaAs QD trion line in top right of (c) is most likely due to the exponential dependence of the spin relaxation rate on the bias close to the co-tunneling region~\cite{Kroner2008}: in the middle of the plateau the highest steady-state fluorescence appears at a bias which maximizes off-resonant pumping and repumping (possible for a small spin splitting), whereas at the edges where spin relaxation is much faster than repumping it appears at a bias corresponding to resonant scattering (similar though less marked features can be seen on all branches on the approach to the co-tunneling zone).
    \textbf{(d)} (\textbf{(e)}) Photon count histogram under time-resolved pulsed resonant excitation at $V=0.26$\,V ($V=0.78$\,V), showing optical spin pumping on the $\ket{\downarrow} \leftrightarrow \ket{\Downarrow \downarrow \uparrow}$ transition, indicated by the red arrow in \textbf{(a)}, at a power above saturation, $\sim 16\mathrm{P_{sat}}$ ($\sim 6\,\mathrm{P_{sat}}$). A decay time of $203(2)\,$ns ($111(1)\,$ns) is obtained from an exponential fit (dashed curve). Insets show the spin pumping rate $\gamma_\text{SP}$ (inverse of decay time) as a function of resonant laser power, where the solid curve is a fit of the data to a saturation model with parameter $\mathrm{P}_\text{sat}$, resulting in $\mathrm{P_{sat}}=7(1)$\,nW ($\mathrm{P_{sat}}=48(4)$\,nW).
    }
    \label{fig1:cyclicity}
\end{figure*}

In this work, we demonstrate quantum control of a QD electron-spin qubit in Faraday geometry -- for both a GaAs QD and an InGaAs QD. 
A small light-hole admixture of the heavy-hole trion weakly breaks the full cyclicity of the spin-selective transitions, while maintaining it at a high value of $471\,(50)$ ($291\,(28)$) for the GaAs (InGaAs) QD, and creates a highly asymmetric $\Lambda$ system. 
Using a narrowband stimulated Raman scheme to drive electron-spin resonance~\cite{Gangloff_quantum_2019}, we account for the GHz-level differential Stark shifts arising from asymmetric Rabi frequencies in the $\Lambda$ system, something otherwise impossible with broadband ps-pulse lasers~\cite{press_complete_2008}. 
With nuclear-spin stabilization of the qubit splitting, we narrow the electron spin resonance to $7(1)\,$MHz ($31(3)\,$MHz), corresponding to an inhomogeneous dephasing time $T_2^*=74(11)\,$ns ($T_2^*=17(2)\,$ns), in a GaAs QD (an InGaAs QD) and achieve phase-tuneable control of the electron-spin qubit -- sufficient for functional spin-photon entanglement~\cite{Meng2024}. With the GaAs QD, we achieve up to $273(6)\,$MHz Rabi frequency and a $\pi$-pulse contrast of $97.4(1)\,\%$. We also perform nuclear-spin spectroscopy and demonstrate the compatibility of this control configuration with a nuclear quantum memory \cite{Appel2025}. Our results demonstrate the technical feasibility of combining high-fidelity coherent control of QD spins and single-shot readout, with direct application to the deterministic generation of large entangled photonic states~\cite{lindner_proposal_2009,Michaels2021}.

Measurements presented in Figs.\,\ref{fig1:cyclicity} are taken on both an InGaAs QD and a GaAs QD. Measurements presented in Figs.\,\ref{fig2:cooling}, \ref{fig3:rabi}, \ref{fig4:ramsey}, and \ref{fig5a:magnon} are taken on a GaAs QD, while a subset of similar measurements performed on an InGaAs QD are presented in Appendix \ref{app:ingaas_ramsey}.
Fig.\,\ref{fig6:comparison} provides a summary on the electron-spin measurements taken on both an InGaAs QD and a GaAs QD.

\section{Devices and Spectroscopy}

We use two different QD devices in this work (Fig.\,1a). 
The first consists of self-assembled InGaAs QDs emitting around $975\,$nm embedded in a Schottky diode (Device 1), while the second consists of droplet-etched GaAs QDs emitting around $775\,$nm embedded in a \textit{p-i-n} diode (Device 2).
Both devices are kept cold in a cryostat at 3.5-4.2~K, and are subject to a magnetic field up to $B_z=8.0\,$T applied along the growth axis ($z$) of the QDs.
Details on the design of both devices can be found in Appendix \ref{app:opticalspin}.

For both devices, we work with a single electron-charged QD, consisting of two Zeeman-split ground electron-spin states $\ket{\downarrow}$ and $\ket{\uparrow}$, and two excited negatively-charged exciton (trion) $X^{1-}$ states $\ket{\Downarrow\downarrow\uparrow}$ and $\ket{\Uparrow\downarrow\uparrow}$.
The trion hole states $\ket{\Uparrow}$ and $ \ket{\Downarrow}$ consist predominantly of heavy-hole (HH) states with a small admixture of light-hole (LH) states~\cite{bayer_fine_2002,koudinov_optical_2004} -- as parametrized by the HH-LH admixture coefficient $\sigma_{lh}/\Delta_{lh} \equiv \epsilon_{lh} \ll 1$, where $\sigma_{lh}$ parametrizes the off-diagonal coupling (in the Luttinger-Kohn Hamiltonian) -- due to strain, disorder, or confinement effects -- between LH and HH states and $\Delta_{lh}$ is the HH-LH energy splitting.
The energy level schematic, depicted in Fig.\,\ref{fig1:cyclicity}(a), highlights the dominant circularly-polarized spin-conserving transitions $\ket{\Downarrow\downarrow\uparrow} \leftrightarrow \ket{\downarrow}$ and $\ket{\Uparrow\downarrow\uparrow} \leftrightarrow \ket{\uparrow}$ and the weaker spin-flipping
transitions, $\ket{\Downarrow\downarrow\uparrow} \leftrightarrow \ket{\uparrow}$ and  $\ket{\Uparrow\downarrow\uparrow} \leftrightarrow \ket{\downarrow}$.  The spin-flipping transitions are forbidden for a pure HH $X^{1-}$ states and weakly enabled in the presence of HH-LH admixture~\cite{belhadj_impact_2010}. Depending on the exact nature of the LH mixing, the spin-flipping transitions can be linear $\pi_z$ -- i.e. along the growth, magnetic-field, and optical axes, and thus mostly dark~\cite{Huo_a_2013} -- or circularly polarized (see Appendix~\ref{app:hhlh}). Despite the large strains present in InGaAs QDs, the spin-conserving HH transitions remain approximately cyclic due to the small QD sizes and thus to a large value of $\Delta_{lh}$ in the tens of meV~\cite{urbaszek_nuclear_2013}. 
While typically larger than $100:1$, the exact branching ratio between the spin-conserving and spin-flipping transitions can vary significantly across multiple QDs in the device due to the strain inhomogeneity of the Stranski–Krastanov growth process~\cite{koudinov_optical_2004,krizhanovskii_individual_2005}.
The origin of the non-zero HH-LH admixture in a typically strain-free GaAs QD can be attributed to 3D confinement effects in nanostructures~\cite{Huo2017} -- a smaller effective $\Delta_{lh}$ of $\sim 3$\,meV together with small geometry- or disorder-induced effective shear strain components at the QD material interface~\cite{Schimpf_Optical_2025}.

Fig.\,\ref{fig1:cyclicity}(b, d) and Fig.\,\ref{fig1:cyclicity}(c, e) present the optical spectroscopy of the $X^{1-}$ transition in an InGaAs QD (Device 1) and a GaAs QD (Device 2) at $B_z=3.5\,$T. A voltage bias $V$ is applied to the Schottky (Device 1) or to the p-i-n (Device 2) diode to enable charge control and tuning of the emission energy via the d.c. Stark shift. Using a dark-field microscope~\cite{kuhlmann_dark_2013}, we scan across the $X^{1-}$ resonance with a linearly polarized narrow-linewidth ($\sim100\,$kHz) laser.  The collected fluorescence from Device 1 (Device 2) is shown in Fig.\,\ref{fig1:cyclicity}(b) (Fig.\,\ref{fig1:cyclicity}(c)) as a function of device bias and laser frequency, at laser powers well above saturation. 
At the edge of the charging plateau (device bias of $\approx 0.22$~V and $\approx 0.30$~V for Device 1 and device bias of $\approx 0.68$~V and $\approx 0.87$~V for Device 2), rapid tunneling of the electron from the QD to the diode back contact reservoir randomizes the spin states~\cite{lu_direct_2010}, thus preventing spin pumping and allowing us to observe steady-state fluorescence on resonance. 
In Device 1, we observe four distinct resonances: two strong and two weak transitions with relative intensities of approximately $100:1$. This is consistent with the level diagram of Fig.\,\ref{fig1:cyclicity}(a), where the two strong (weak) transitions correspond to the spin-conserving (flipping) transitions,
and demonstrates that it is possible to directly excite the normally forbidden spin-flipping transitions. We further extract an electron-spin splitting of $30$\,GHz and a trion linewidth of $471(36)$\,MHz from a Lorentzian fit to all four peaks at this bias. 
For Device 2, we observe only two sharp resonances, corresponding to the spin-conserving transitions; this is likely due to the small electron-spin splitting -- a few GHz for GaAs QDs at this wavelength and field~\cite{Schimpf_Optical_2025} -- making the direct excitation of weak spin-flipping transitions difficult to resolve in close proximity to the strong spin-conserving transitions. The small spin splitting of the GaAs QD is also responsible for the reduced contrast, relative to the InGaAs QD, between the edges and the center of the charging plateau (see Appendix~\ref{app:opticalspin}). The splitting between the two spin-conserving transitions reveals a combined electron and hole splitting of $\omega_h +\omega_e=81.6(2)\,$GHz. For this QD, we measure a trion linewidth of $851(27)\,$MHz from a Lorentzian fit to the data taken at a device bias of 0.75~V and at zero field. The discontinuity in the GaAs QD trion line in top right of Fig.~\ref{fig1:cyclicity}(c) is most likely due to the exponential dependence of the spin relaxation rate on the bias close to the co-tunneling region~\cite{Kroner2008}: in the middle of the plateau the highest steady-state fluorescence appears at a bias which maximizes off-resonant pumping and repumping (possible for a small spin splitting), whereas at the edges where spin relaxation is much faster than repumping it appears at a bias corresponding to resonant scattering (similar though less marked features can be seen on all branches on the approach to the co-tunneling zone).

Focusing now on a device bias in the middle of the charging plateau, we observe a significantly lower intensity from the dominant peaks compared to that at the edge of the plateau (Fig.\,\ref{fig1:cyclicity}(b,c)). Despite the weakness of the spin flipping process in this field configuration (allowed by HH-LH mixing~\cite{Schimpf_Optical_2025}), this measurement confirms the presence of significant optical pumping of the electron-spin population against background spin relaxation mechanisms. A measurement of spin relaxation time $T_1$, exceeding $40$\,$\mu$s for both devices, is presented in Appendix\,\ref{app:opticalspin}.
We then probe the time scale of the spin-pumping process by measuring the time it takes for the electron spin to initialize to $\ket{\uparrow}$ by continuously driving the $\ket{\downarrow} \leftrightarrow \ket{\Downarrow \downarrow \uparrow}$ transition for $5\,\mu s$, shown in Fig.\,\ref{fig1:cyclicity}(d) and Fig.\,\ref{fig1:cyclicity}(e) for the two devices, respectively. 
Taken with a readout power of $16$ ($6$) times the saturation power $P_\text{sat}$, to approach the minimum pumping time, our measurement reveals a fitted exponential decay time of $\gamma_\text{SP}^{-1}=203(2)\,$ns ($\gamma_\text{SP}^{-1}=111(1)\,$ns) for Device 1 (2), yielding the maximum spin-flipping scattering rate $\gamma_\text{SP}$. The steady-state fluorescence at long delay time is dominated by off-resonant repumping of the spin state by the readout laser at these high powers. For this reason, in subsequent measurements presented in this article for both Devices 1 and 2, we perform spin initialisation at lower powers of $5P_{\rm sat}$ and $0.5P_{\rm sat}$ (and for Device 2 at a higher magnetic field) for which repumping is no longer significant and an initialisation fidelity upper bound of $98\%$ and $97\%$ can be reached, respectively (as detailed in Appendix\,\ref{app:opticalspin}).
Using the previously reported $X^{1-}$ lifetime of $\gamma_1^{-1}= 0.696\,(65)\,$ns~\cite{Stanley_dynamics_2014,stockill_phase-tuned_2017} for an InGaAs QD on this device and an estimated lifetime of $\gamma_1^{-1}=0.235\,(25)$~ns (c.f. Appendix~\ref{app:lifetime}) for the GaAs QD, we obtain an optical cyclicity of $\mathcal{C} \equiv \gamma_\mathrm{SC}/\gamma_\mathrm{SP} = \gamma_1/\gamma_\mathrm{SP} - 1 = 291\,(28)$ and $471\,(50)$, where $\gamma_\mathrm{SC}$ is the spin-conserving scattering rate, and a branching ratio for the spin-flipping transition $\gamma_\text{SP}/\gamma_1$~\cite{dreiser_optical_2008,lu_direct_2010} of 0.34\,(3)$\%$ and 0.21\,(2)\,$\%$, respectively. 
The cyclicity $\mathcal{C}$ determines the mean number of photons scattered by the bright spin state before it flips to the dark spin state during a readout event, and sets the upper limit for the single-shot readout fidelity~\cite{rosenthal_single_2024}.

\section{Stimulated Raman Transitions in Faraday Configuration}

\begin{figure*}
    \centering 
    \includegraphics[width=0.75\textwidth]{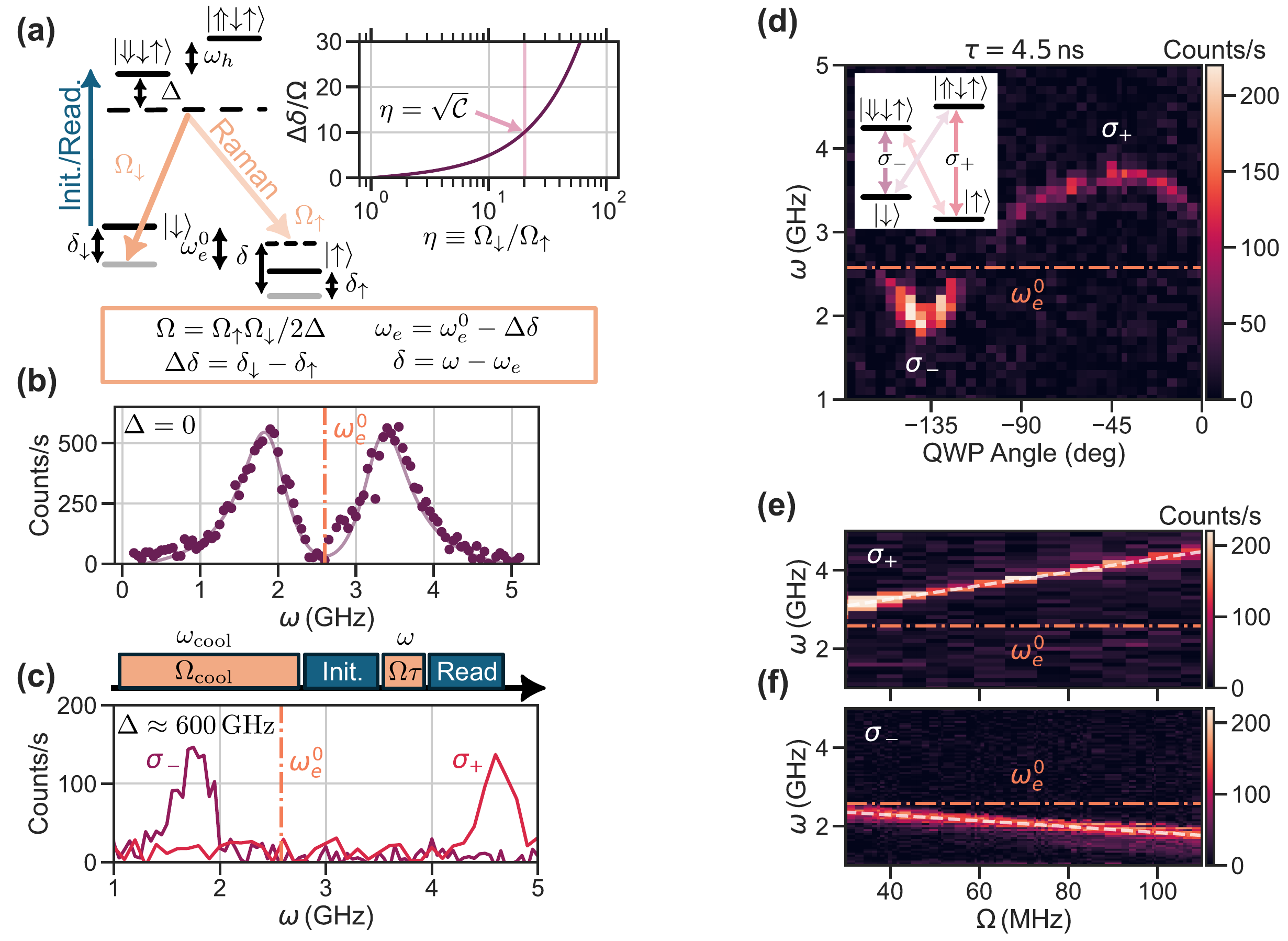}
    \caption{
    \textbf{Dependence of electron spin resonance on Raman laser at $B_z=6.5\,$T.} All measurements are taken on a GaAs QD in Device 2.
    \textbf{(a)} Energy levels relevant for Raman spin control. Each Raman field exerts an a.c. Stark shift (gray lines) $\delta_\downarrow$ and $\delta_\uparrow$ on its respective ground states. The right panel shows $\Delta \delta/\Omega$ as a function of $\eta = \Omega_\downarrow/\Omega_\uparrow$, as related to cyclicity by $\eta=\sqrt{\mathcal{C}}$.
    \textbf{(b)} Continuous two-color resonant ($\Delta=0$) measurement with a fixed-frequency laser (pump) set to set to $700\,$nW ($\approx 14\, \mathrm{P_{sat}}$) and a variable frequency laser (probe) set to $70$\,nW ($\approx 1.4\, \mathrm{P_{sat}}$). The resulting fluorescence signal is shown as a function of relative pump-probe (two-photon) detuning $\omega$. A fit (solid curve) to the data with a three-level model (see Appendix \ref{app:cpt}) reveals an electron-spin splitting of $\omega_e^0 = 2.60\,(1)\,$GHz from the dark-state (dip) resonance condition $\omega = \omega_e^0$.
    \textbf{(c)} Top: pulse sequence for obtaining an ESR spectrum consisting of nuclear-spin cooling ($40\,\mu s$ of Rabi cooling) at Rabi frequency $\Omega_\text{cool}=\Omega$ and drive frequency $\omega_\text{cool}=\omega$ followed by $1.5\,\mu s$ of spin initialization, a short probe with Rabi frequency $\Omega$, drive frequency $\omega$ and duration $\tau$, and a $1.5\,\mu s$ spin readout. Bottom: readout fluorescence as a function of drive frequency. Measurements are performed at $\Omega \approx 110$\,MHz and $\tau = 4.5$\,ns using the Raman laser polarization settings aligned to $\sigma_-$ (purple data) and $\sigma_+$ (red data).
    \textbf{(d)} ESR spectrum as a function of Raman laser polarization set by the quarter-wave-plate (QWP) angle. QWP angle of $0^\circ$ is calibrated to correspond to linear horizontal polarization, and the right-hand ($\sigma_+$) and left-hand ($\sigma_-$) circular polarizations are at $-45^\circ$ and $-135^\circ$, respectively. 
    \textbf{(e)} (\textbf{(f)}) ESR spectrum as a function of Rabi frequency for right-hand (left-hand) circular polarization. Dashed lines are linear fits to the frequency corresponding to a fluorescence maximum, with fitted slopes $\Delta \delta/\Omega = -17.2\,(6)$ and $\Delta \delta/\Omega = 7.4\,(5)$.
    Both fits converge to $\omega_e^0=2.58\,(3)\,$GHz at the limit of vanishing $\Omega$.
    }
    \label{fig2:cooling}
\end{figure*}

Figure~\ref{fig2:cooling}(a) shows the relevant energy levels involved in our spin control experiment. For readout, a resonant laser drives the lowest energy cycling transition $\ket{\Downarrow\downarrow\uparrow} \leftrightarrow \ket{\downarrow}$ at $~0.5\,\mathrm{P_{sat}} (\approx 25\,$nW) for both initialization to the spin $\ket{\uparrow}$ state and readout of the $\ket{\downarrow}$ state.
To coherently control the ground state spin, we use an all-optical Raman drive that is circularly polarized and red-detuned by the single-photon detuning $\Delta \approx 600\,$GHz from the $\ket{\Downarrow\downarrow\uparrow}$ trion. The two Raman fields are derived from microwave-driven electro-optical modulation of a single narrow-band continuous-wave laser, resulting in a coherent pair of fields $\vec{E}_{\downarrow}$ and $\vec{E}_{\uparrow}$ equal in both intensity and polarization~\cite{Bodey_Opitcal_2019}. The two Raman fields, with resonant Rabi frequencies $\Omega_{\downarrow}$ and $\Omega_{\uparrow}$, respectively, induce a rotation of the ground-state spin at a two-photon Rabi frequency $\Omega = \Omega_{\downarrow}\Omega_{\uparrow}/2\Delta$ when their frequency difference, the two-photon detuning $\omega$, matches the ground-state splitting $\omega_e$, i.e. achieving electron-spin resonance (ESR) at $\delta \equiv \omega - \omega_e = 0$. The small branching ratio of the spin-flipping transition implies a large imbalance in the Rabi frequencies of the two transitions -- $\Omega_{\downarrow} \gg \Omega_{\uparrow}$ ($\Omega_{\downarrow} \ll \Omega_{\uparrow}$) for left-circularly (right-circularly) polarized Raman beams -- which results in a large differential (a.c.) Stark shift $\Delta \delta \equiv \delta_{\downarrow} - \delta_{\uparrow} = (\Omega^2_{\downarrow}-\Omega^2_{\uparrow})/4\Delta$ on the electron-spin states during the Raman drive. We thus tune $\omega$ to match the ground state spin splitting $\omega_e = \omega_e^0 - \Delta \delta$, where $\omega_e^0$ is the bare electron-spin splitting, and probe Raman resonance ($\omega \sim  \omega_e$) when driven.

To visualize the effect of this imbalance, we can express the two-photon Rabi frequency $\Omega$ and the differential Stark shift $\Delta \delta$ in terms of $\Omega_\downarrow$ and the Rabi frequency imbalance, $\eta \equiv \Omega_{\downarrow}/\Omega_{\uparrow}$ as 
\begin{equation}\label{eqn:rabi_freq}
    \Omega =\frac{\Omega_\downarrow\Omega_\uparrow}{2 \Delta}=\frac{\Omega_\downarrow^2}{2\eta\Delta} \\ 
\end{equation}
and
\begin{equation}\label{eqn:light_shift}
    \Delta \delta = \frac{\Omega_\downarrow^2 - \Omega_\uparrow^2}{4\Delta}= \Omega \frac{\eta^2-1}{2\eta},
\end{equation}
respectively. 
In the situation where the two Raman fields have equal intensity, $|\vec{E}_{\downarrow}|^2 = |\vec{E}_{\uparrow}|^2$, and are exactly matched to two transition dipoles $\vec{d}_{\downarrow}$ and $\vec{d}_{\uparrow}$ (i.e. the expectation value of the electric dipole Hamiltonian $\langle\vec{d}_\cdot\vec{E}\rangle$ is maximized),

\begin{align}\label{eqn:matched}
    \eta^2 &= \left(\frac{\Omega_{\downarrow}}{\Omega_{\uparrow}}\right)^2 
      = \left(\frac{\langle\vec{d}_{\downarrow}\cdot\vec{E}_{\downarrow}\rangle}
                   {\langle\vec{d}_{\uparrow}\cdot\vec{E}_{\uparrow}\rangle}\right)^2 \notag \\
    &\rightarrow \left(\frac{|\vec{d}_{\downarrow}|}{|\vec{d}_{\uparrow}|}\right)^2 
      = \frac{\gamma_\text{SC}}{\gamma_\text{SP}} 
      = \mathcal{C},
\end{align}

\noindent(or $=\gamma_\text{SP}/\gamma_\text{SC} = \mathcal{C}^{-1}$) for left-circular (right-circular) polarization. Using our measured cyclicity of $\mathcal{C}=471\,(50)$, we can thus expect a shift in the ESR resonance by $11(1) \,\Omega$. At a typical Rabi frequency of $\Omega=100\,$MHz, this corresponds to shifts of $|\Delta\delta|\simeq 1\,$GHz, which exceeds even the inhomogeneous ESR linewidth ($\sim 300\,$MHz~\cite{Gangloff_quantum_2019, Zaporski_Ideal_2023, nguyen_enhanced_2023}).

As we are unable to resolve the electron splitting via spectroscopy for the GaAs QD in Device 2 (Fig.\,\ref{fig1:cyclicity}(c)), we perform two-laser resonant spectroscopy at a fixed bias of 0.775\,V and a magnetic field of 6.5\,T with a fixed pump laser frequency at 386.435\,THz while varying the frequency of a second probe laser. The pump and probe laser powers are fixed at approximately $ 14\, \mathrm{P_{sat}}$ and $ 1.4\,\mathrm{P_{sat}}$, respectively. The resulting steady-state fluorescence data is shown in  Fig.\,\ref{fig2:cooling}(b) and shows a dip in the signal, a signature of coherent population trapping (CPT)~\cite{Xu2008,Brunner2009,ethier_2017}, when the two lasers meet the two-photon resonance condition for the spin-conserving and spin-flipping transitions. Following a fit to the measurement data using the model and parameters described in Appendix~\ref{app:cpt}, the dip minimum reveals a bare electron splitting of $\omega_e^0=2.60(1)\,$GHz -- corresponding to an out-of-plane $g$-factor of $0.0286(1)$, as consistent with previous measurements~\cite{Schimpf_Optical_2025}. 

Finding the ESR frequency in a Raman experiment typically requires stabilizing the Overhauser shift imparted by nuclear spins on the electron-spin splitting~\cite{Gangloff_quantum_2019}. We do this using nuclear-spin cooling techniques -- one of Raman~\cite{Gangloff_quantum_2019}, Rabi~\cite{nguyen_enhanced_2023}, or quantum-algorithmic~\cite{jackson_optimal_2022} cooling, depending on the experiment.
Our typical pulse sequence, shown in Fig.\,\ref{fig2:cooling}(c), consists of a cooling period  -- during which our coherent Raman drive and resonant readout serve to sense and correct the nuclear polarization -- followed by a probe experiment (representing $\sim 10\%$ of the duty cycle) in which the electron spin is initialized, driven coherently by Raman beams for a fixed duration, and read out.  
Fixing the probe Rabi frequency to $\Omega\approx110$~MHz and the pulse duration to $\tau=1/(2\Omega)=4.5\,$ns, i.e. $\pi$ pulse, we measure the readout fluorescence as a function of probe frequency $\omega$ -- here the Raman cooling frequency $\omega_\text{cool}$ is also set to $\omega$. We observe a clear electron-spin resonance (ESR) signal shifted from $\omega_e^0$ by about $\sim 1$\,GHz ($2$\,GHz) down (up) for left-circularly (right-circularly) polarized Raman light. The sign of the shift matches our expectation as a red-detuned beam lowers the energy of the ground state; here, the stronger spin-conserving left-circularly (right-circularly) polarized transition lowers the $\downarrow$ ($\uparrow$) state more than the spin-flipping transition lowers the state $\uparrow$ ($\downarrow$), resulting in a decrease (an increase) in the electron-spin splitting that is proportional to the Raman laser intensity. 

Fig.~\ref{fig2:cooling}(d) shows a summary of this measurement as a function of the angle of the quarter-wave plate (QWP) used to tune the circularity of the Raman laser, where $0^o$ is calibrated to correspond to linear horizontal polarization. We observe clearly that the Raman laser must exhibit a well-defined handedness for ESR to be possible. Moreover, the maximal amplitude and maximally-positive and negative shifts in the ESR resonance occur at $-45^o$ and $-135^o$, respectively, corresponding to $\sigma_+$ and $\sigma_-$ polarization settings and matching the dipoles of the two spin-conserving transitions (further details in Appendix~\ref{app:polarization}). This implies destructive interference of the two $\Lambda$ systems when driving with linearly polarized light, which is expected for a specific kind of HH-LH mixing as shown in Appendix~\ref{app:hhlh} -- by contrast, on the InGaAs QD in Device 1 a degree of linear diagonal polarization seems to improve the Raman coupling (see Appendix \ref{app:ingaas_ramsey}), which implies a different kind of HH-LH mixing (see Appendix~\ref{app:hhlh}). We further observe a larger maximal shift from the undriven resonance frequency $\omega_e^0$ (horizontal dashed line) for $\sigma_+$ than for $\sigma_-$, consistent with the difference in ESR shifts in Fig.~\ref{fig2:cooling}(c), and a clear deviation from a simple sinusoidal dependence on the QWP angle. While we expect a small difference in coupling between $\sigma_+$ and $\sigma_-$ polarizations, due to the trion splitting making the Raman detuning for $\sigma_+$ larger by $\omega_h \approx 150$\,GHz, this asymmetric response to $\sigma_+$ and $\sigma_-$ suggests an additional mechanism is present.

Further evidence for this is provided by Fig.~\ref{fig2:cooling}(e,f), which shows a summary of ESR measurements now taken as a function of probe Rabi frequency $\Omega$, for right-circular (e) and left-circular polarization (f). We observe a clear linear dependence on $\Omega$ of the ESR peak frequency $\omega_e = \omega_e^0 - \Delta \delta$, as expected from Eqn.\,\ref{eqn:light_shift}. Fitting $\omega_e$ as a function of $\Omega$ with two independent linear functions, we obtain $\Delta\delta/\Omega=-17.2(6)$ and $\Delta\delta/\Omega=7.4(5)$, corresponding to Rabi frequency imbalances of $\eta^{-1}=34(1)$ and $\eta = 15(1)$, respectively, and an undriven resonance frequency $\omega_e^0 = 2.58(3)$\,GHz that agrees closely between the two linear fits (and is within one standard error of our CPT result in Fig.\,\ref{fig2:cooling}(b)). These values of $\eta$ deviate measurably from the idealized value derived from our measured optical cyclicity $\eta = \sqrt{\mathcal{C}} = 22(1)$. Equations~\ref{eqn:rabi_freq} and \ref{eqn:light_shift} tell us that the differential Stark shift is dominated by the spin-conserving transition Rabi frequency $\Omega_\downarrow$ ($\Omega_\uparrow$) for $\sigma_-$ ($\sigma_+$), whereas the two-photon Rabi frequency $\Omega$ is equally affected by the spin-conserving and spin-flipping contributions. The unequal response of the ESR frequency as a function of Rabi frequency between $\sigma_+$ and $\sigma_-$ polarizations thus
reveals that the Raman beam coupling to the spin-flipping transition must differ significantly between the two polarization settings.

We propose the following explanation for this asymmetric response to $\sigma_+$ and $\sigma_-$, as further detailed in Appendix~\ref{app:hhlh}. In general, the LH admixtures to the $\ket{\Uparrow}$ and $\ket{\Downarrow}$ are not equal in a magnetic field, an effect which is more pronounced for GaAs QDs due to a lower value of $\Delta_{lh}$. The kinetics of the two lambda systems are thus asymmetric, rendering $\eta$ and $\mathcal{C}$ dependent on the helicity and magnetic-field strength. A simple evaluation of the hole-spin-dependent value of $\Delta_{lh}$ in a 6.5\,T field, using known GaAs values for the LH and HH g-factors, yields a predicted ratio between $\eta^{-1}$ for $\sigma_+$ and $\eta$ for $\sigma_-$ of $\sim 2$, which agrees well with our measured ratio of $2.3(2)$. This asymmetry reveals a potentially useful feature for GaAs QDs at high magnetic fields: that there is a transition with enhanced cyclicity -- here $\sigma_+$ with $\mathcal{C}=\eta^{-2}\sim 1100$ -- best suited for readout, and another with reduced cyclicity -- here $\sigma_-$ with $\mathcal{C}=\eta^{2}\sim 225$ -- best suited for spin control and spin initialization.

The results in Fig.\,\ref{fig2:cooling} highlight the essence of optical spin control in the Faraday configuration: an inevitable shift in the ESR frequency that depends on Raman laser polarization and the instantaneous Rabi frequency. If a $\pi_z$ component is present in the spin-flipping transition (as we suppose is the case for the InGaAs QD but not the GaAs QD), then this shift could be mitigated, though not eliminated, with an asymmetric Raman drive. In all instances, it must be calibrated, as we have done, before experiments with programmable pulse amplitude and phase can be performed.

\begin{figure*}
    \centering
    \includegraphics[width=1\textwidth]{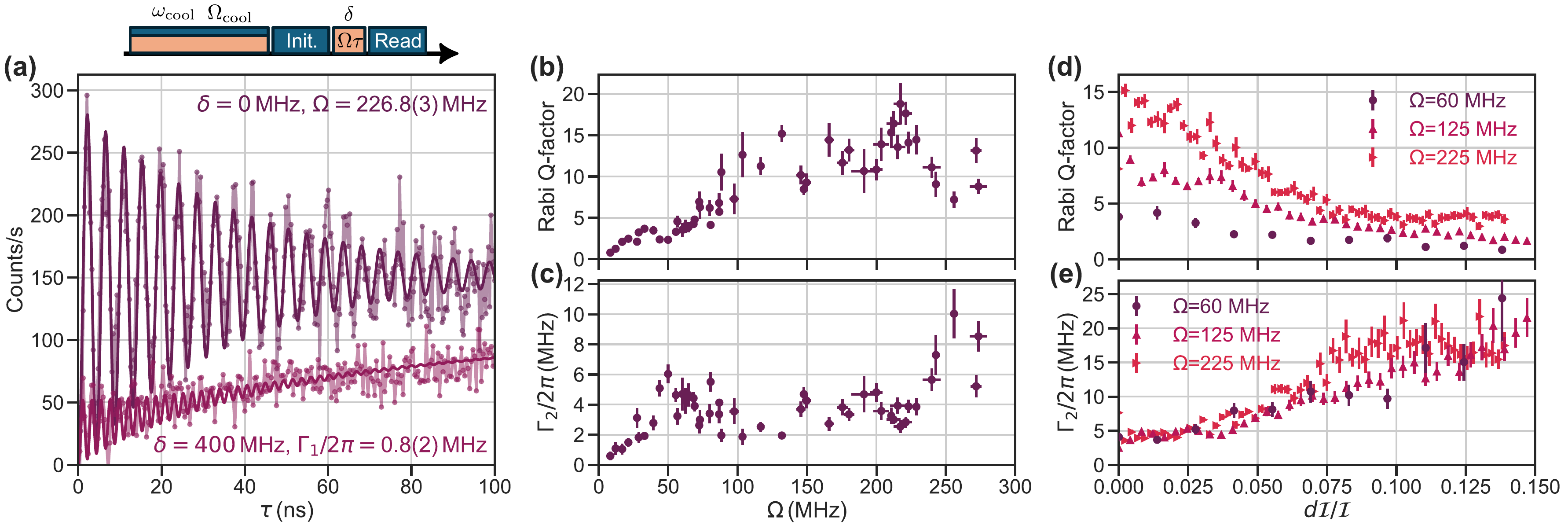}
    \caption{
    \textbf{Coherent control of the electron spin.} All measurements are taken on a GaAs QD in Device 2.
    \textbf{(a)} Measurement of Rabi oscillations with pulse sequence (top) consisting of nuclear-spin cooling (Raman cooling) at Rabi frequency $\Omega_\text{cool}$ and drive frequency $\omega_\text{cool}$ followed by $1.5\,\mu s$ of spin initialization, a probe pulse with Rabi frequency $\Omega$, detuning $\delta$ and duration $\tau$, and a $1.5\,\mu s$ of spin readout. Measurement data is readout counts as a function of $\tau$, and shows Rabi oscillations at $\delta=0$ and laser-induced spin relaxation at $\delta = 400$\,MHz. The solid curves are a master equation model (Appendix \ref{app:model}), with fixed $T_2^*=34$\,ns, used to fit the relaxation rate $\Gamma_1/2\pi = 0.8(2)$\,MHz at $\delta = 400$\,MHz and the Rabi frequency $226.8(3)\,$MHz and dephasing rate $\Gamma_2/2\pi = 3.7(1)$\,MHz at $\delta = 0$\,MHz. The Rabi quality factor is $Q = 15.1(6)$. 
    \textbf{(b)} Rabi quality factor $Q$ and \textbf{(c)} spin dephasing rate $\Gamma_2/2\pi$ as a function of $\Omega$ for $\Omega_\text{cool}=15$\,MHz. The optical power of the Raman laser entering the cryostat for $\Omega=250\,$MHz is approximately $290\, \mu\mathrm{W}$.  
    \textbf{(d)} Rabi quality factor $Q$ and \textbf{(e)} spin dephasing rate $\Gamma_2/2\pi$ as a function of artificially-introduced Raman laser intensity modulation $d\mathcal{I}/\mathcal{I}$ at three probe Rabi frequencies $\Omega$. 
    Error bars indicate one standard error on fitted parameters.
    }
    \label{fig3:rabi}
\end{figure*}

\section{Rabi Oscillations and Quality Factor}

With the stabilization of the ESR and the calibration of the differential Stark shift achieved, we proceed to show time-resolved coherent driving of the electron spin on resonance. 
For that, we measure the readout counts as a function of probe pulse duration following stabilization of the nuclei via Raman cooling, as depicted in Fig.\,~\ref{fig3:rabi}(a). 
The resulting signal shows damped Rabi oscillations when the ESR drive is resonant ($\delta=0$). 
When the detuning is increased to $\delta = 400\,$MHz, we observe a monotonic increase in bright-state population once a few detuned Rabi oscillations have decohered ($\tau \gtrsim 30$\,ns); an exponential fit yields a relaxation rate of $\Gamma_1/2\pi=0.8(2)\,$MHz and provides a direct measurement of background optically induced spin relaxation~\cite{Bodey_Opitcal_2019}. 
The purple solid curve is a fit to a master equation model (see Appendix~\ref{app:model}) accounting for the measured inhomogeneous broadening of $14\,$MHz ($T_2^*=34$\,ns), the measured spin relaxation rate $\Gamma_1$, and spin dephasing with a free parameter $\Gamma_2$. 
From this fit, we extract a Rabi frequency of $\Omega=226.8(3)\,$MHz (or a $\pi$-time of $t_\pi=1/2\Omega=2.195(2)\,$ns) and spin dephasing rate of $\Gamma_2/2\pi = 3.7(1)\,$MHz. From the fitted density matrix, we also extract a $\pi$-pulse contrast of $96.8(1)\%$, which is an upper bound on the $\pi$-pulse fidelity~\cite{Bodey_Opitcal_2019}, and a corresponding Rabi oscillation quality factor $Q=-1/\ln(2f_\pi-1) = 15.1(6)$. 

Fig.\,\ref{fig3:rabi}(b) and (c) shows the quality factor $Q$ and spin dephasing rate $\Gamma_2$, extracted from our master equation model as in Fig.\,\ref{fig3:rabi}(a), as a function of probe Rabi frequency $\Omega$. 
% To account for the finite inhomogeneous broadening (quantified by the $T_2^*$ value), we would first fit the data with our model (that takes into account of the spin relaxation, dephasing and inhomogeneous broadening effects), estimate the $\pi$-pulse fidelity from the fit $f_\pi$, which we can then convert to $Q$ via $f_\pi = 0.5 \left(1-\exp(-1/Q)\right)$.
Our optimal factor of $Q=18.8(3)$, corresponding to a $\pi$-pulse contrast of $97.4(1)\%$, is obtained for $\Omega=217(5)\,$MHz. The overall increase in $Q$ towards this value is expected from previous work~\cite{Bodey_Opitcal_2019,nguyen_enhanced_2023}, where Rabi frequencies well beyond $(2\pi T_2^*)^{-1}$ and nuclear Zeeman frequencies ($50-80$\,MHz) enable decoupling from nuclear effects. This behavior is also reflected in our fitted $\Gamma_2$ values, which start out low due to inhomogeneity-dominated dephasing at low $\Omega$, peak at $4-6$\,MHz around $\Omega=50$\,MHz (i.e. the Arsenic nuclear Larmor frequency) and taper off to $2-4$\,MHz at higher $\Omega$ values. The subsequent decrease in $Q$ and increase in $\Gamma_2$ for $\Omega \gtrsim 230$\,MHz, however, is likely due to a mechanism made worse by Faraday-geometry spin control: laser intensity noise. Any fluctuation in Raman laser intensity not only changes the instantaneous Rabi frequency but also the electron-spin splitting directly via the differential Stark shift. As $\Omega$ scales linearly with laser intensity, $\mathcal{I}$, we can determine the fluctuation in the ESR shift as $d(\Delta \delta)/\Delta \delta= d\mathcal{I}/\mathcal{I}$. In the experiment, we measured a typical laser power fluctuation $ \sim
1\,\%$ (integrated over the full bandwidth of our photo-detector) that worsened at our maximum laser intensities. With a differential Stark shift of $\Delta\delta=7.4(5)\, \Omega$ (see Fig.\,\ref{fig2:cooling}(f)), $\Omega=250\,$MHz would result in a broadening of the ESR resonance $d(\Delta \delta) \lesssim 18.5\,$MHz. While our fitted dephasing rate of $\Gamma_2/2\pi = 8-10$\,MHz for $\Omega \approx 250$\,MHz is significantly lower than this bound, this could be explained by a partial decoupling from the noise spectrum below $\Omega$ in a resonant Rabi experiment~\cite{Bodey_Opitcal_2019}. To verify this further, we artificially introduce broadband intensity noise (up to $35$\,MHz) on the Raman laser and measure the resulting Rabi oscillations. The summary is shown in Fig.\,\ref{fig3:rabi}(d) and (e), where a steady decrease of $Q$ and a steady increase of $\Gamma_2$, respectively, as a function of $d\mathcal{I}/\mathcal{I}$ can be observed for $d\mathcal{I}/\mathcal{I} \gtrsim 0.025$. 
These behaviors are qualitatively matched by our master equation model (see Appendix~\ref{app:intensitynoise}), which treats the added intensity noise as a source of inhomogeneity.

\begin{figure*}
    \centering
    \includegraphics[width=1\textwidth]{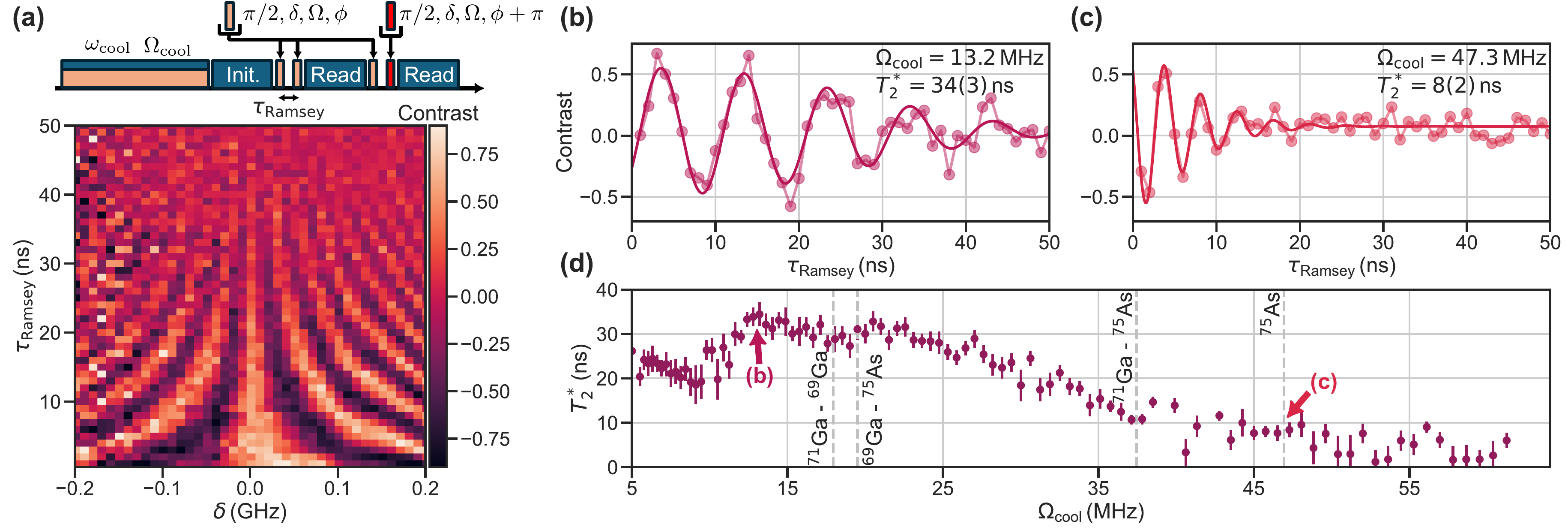}
    \caption{
    \textbf{Phase-resolved control of the electron-spin qubit.} All measurements are taken on a GaAs QD in Device 2.
    \textbf{(a)} Ramsey measurement with pulse sequence (top) consisting of nuclear-spin cooling ($40\,\mu s$ of Raman cooling) at Rabi frequency $\Omega_\text{cool}$ and drive frequency $\omega_\text{cool}$ followed by two Ramsey sequences. The first Ramsey sequence consists of $1.5\,\mu s$ of spin initialization, a first $\pi/2$ rotation with phase $\phi$ (i.e. our phase reference), a Ramsey delay time $\tau_\text{Ramsey}$, a second $\pi/2$ rotation with phase $\phi$, and a $1.5\,\mu s$ of spin readout with average count $n_\phi$. The second Ramsey sequence consists of spin initialization, a first $\pi/2$ rotation with phase $\phi$, a Ramsey delay time $\tau_\text{Ramsey}$, a second $\pi/2$ rotation with phase $\phi + \pi$, and a spin readout with average count $n_{\phi+\pi}$. Our signal is the contrast $(n_\phi-n_{\phi+\pi})/(n_\phi+n_{\phi+\pi})$. All rotations are performed with $\Omega=125\,$MHz and a detuning $\delta$.  Measurement data shows contrast as a function of $\delta$ and $\tau_\text{Ramsey}$ for $\Omega_\text{cool}=15$\,MHz.
    \textbf{(b)} (\textbf{(c)}) Contrast as a function of $\tau_\text{Ramsey}$ for $\Omega_\text{cool}=13.2$\,MHz ($\Omega_\text{cool}=47.3$\,MHz). The solid curve is a fitting function $A\sin(2\pi\delta\tau_\text{Ramsey}+\theta)\exp(-(\tau_\text{Ramsey}/T_2^*)^2)$ with free parameters: $A$ as a contrast amplitude, $\delta$ as the detuning from ESR, $\theta$ as a phase, and $T_2^*$ as the characteristic decay time. 
    The relevant fit results are $T_2^* = 34(3)$\,ns ($T_2^* = 8(2)$\,ns) and $\delta = 100(1)\,$MHz ($\delta = 229(4)\,$MHz).
    \textbf{(d)} Inhomogeneous dephasing time $T_2^*$, fitted from Ramsey experiments as performed in (b,c), as a function of $\Omega_\text{cool}$. Dashed lines highlight expected single- and double-species nuclear resonances within this range of frequencies. Error bars indicate one standard error on fitted parameters.
    }
    \label{fig4:ramsey}
\end{figure*}
\section{Quantum Control and Coherence}

Next we demonstrate complete control over the electron-spin qubit by performing phase-modulated Ramsey interferometry. Our sequence, shown in Fig.\,\ref{fig4:ramsey}(a), consists of two sequences, one with and one without an extra $\pi$ phase on the second $\pi/2$ pulse, which follows a delay time $\tau_\text{Ramsey}$ from a first $\pi/2$ pulse. This two-sequence configuration thus reads out both spin states at every delay time, providing a balanced readout. The readout counts are given by $n_\phi$ and $n_{\phi+\pi}$, respectively, where $\phi$ is the pulse phase, and our signal is then the contrast $(n_\phi - n_{\phi+\pi})/(n_\phi + n_{\phi+\pi})$. Fig.~\ref{fig4:ramsey}(a) shows the Ramsey contrast as a function of drive detuning $\delta$ and Ramsey delay $\tau_\text{Ramsey}$, displaying the expected chevron pattern in the presence of sufficient coherence to observe clearly resolved fringes. 

To measure the inhomogeneous dephasing time $T_2^*$ we take the Ramsey data at a finite detuning $\delta$ from ESR, corresponding to a Ramsey-delay-dependent phase $\phi + 2\pi \delta\tau_\text{Ramsey}$ on the second $\pi/2$ pulse ($\phi$ as our phase reference for the first pulse), and thus contrast oscillations at a frequency $\delta$ to enable better fitting to the decay envelope. This is shown in Fig.\,\ref{fig4:ramsey}(b), where we extract $T_2^* = 34(3)\,$ns from a Gaussian envelope fit with $\delta=100(1)\,$MHz. As known from previous work~\cite{Gangloff_quantum_2019}, the dephasing time $T_2^*$ depends on the cooling Rabi frequency $\Omega_\text{cool}$ -- as shown by a striking decrease to $T_2^*=8(2)\,$ns for $\Omega_\text{cool}=47.3$\,MHz (Fig.\,\ref{fig4:ramsey}(c)). A full summary of the dependence of $T_2^*$ on $\Omega_\text{cool}$ is shown in Fig.\,\ref{fig4:ramsey}(d), where the optimal cooling performance is found for $\Omega_\text{cool}\sim 15$\,MHz. This is consistent with previous results in the Raman cooling configuration, where the optimal cooling Rabi frequency was a fraction of the nuclear Zeeman frequency~\cite{Gangloff_quantum_2019}.

\section{Origins of nuclear cooling}

\begin{figure*}
    \centering    \includegraphics[width=1\textwidth]{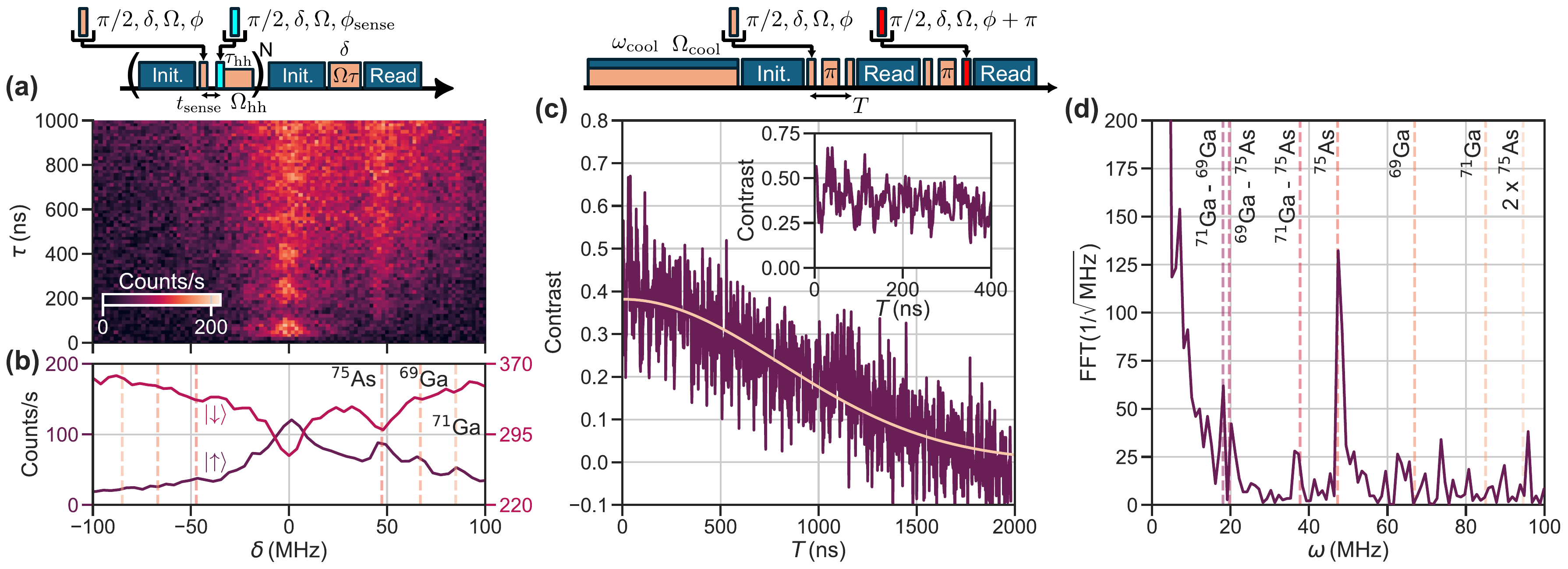}
    \caption{\textbf{Nuclear-magnon and Hahn-Echo spectroscopy}. All measurements are taken on a GaAs QD in Device 2.
    \textbf{(a)} Nuclear-magnon spectroscopy measurement with pulse sequence (top panel) consisting of nuclear-spin cooling ($55\,\mu s$ of modified quantum algorithmic cooling; see Appendix~\ref{modified_ramsey_cooling}) at Rabi frequency $\Omega_\text{hh}=47$\,MHz followed by a probe Rabi sequence consisting of $1.5\,\mu s$ of spin initialization, a Rabi drive at Rabi frequency $\Omega = 6$\,MHz, detuning $\delta$, and of duration $\tau$, and $1.5\,\mu s$ of spin readout. The color map shows the resulting average readout counts as a function of $\delta$ and $\tau$. 
    \textbf{(b)} Spectrum, acquired as in (a), averaged over all Rabi drive durations when the electron spin is initialized to $\ket{\uparrow}$ (purple), as in (a), or to $\ket{\downarrow}$ (pink), with an additional $\pi$ pulse before the Rabi drive; the readout remains on the $\ket{\downarrow}$ state.  
    \textbf{(c)} Hahn Echo spectroscopy measurement with pulse sequence (top) consisting of the same pulse sequence as in Fig.\,\ref{fig4:ramsey}a with the addition of refocusing $\pi$ pulses halfway between the Ramsey pulses. The resulting contrast as a function of total delay time $T$ is shown below. The solid curve is a fit to the function $A\exp\left(-(T/T_2^\text{HE})^2\right)$, where $T_2^\text{HE}=1.14(2)$\,$\mu$s and $A$ is the amplitude of the envelope. Inset: zoom in of the same data. \textbf{(d)} Fourier transform of the data in (c). Dashed lines are expected nuclear Larmor frequencies. Data in (c) and (d) is taken at a cooling frequency $\omega_\mathrm{cool}=2.3$ \,GHz.
    } 
    \label{fig5a:magnon}
\end{figure*}

To probe the nature of the electro-nuclear interaction, we run nuclear-magnon spectroscopy~\cite{Gangloff_quantum_2019}, which consists of a Rabi probe sequence where the drive detuning and duration are varied, while the drive Rabi frequency is fixed to $\Omega =6$\,MHz. The results are shown in Fig.\,\ref{fig5a:magnon}(a) and Fig.\,\ref{fig5a:magnon}(b), where at $\delta\approx0$ we observe our typical ESR signal and Rabi oscillations, with a full-width at half-maximum of $7(1)$\,MHz consistent with $T_2^* = 74(11)$\,ns  -- obtained with modified quantum algorithmic cooling (see Appendix~\ref{modified_ramsey_cooling}). At a finite detuning $\delta \approx 47$\,MHz, which matches the Arsenic (As) Larmor frequency, we see another resonance appear, corresponding to a magnon mode of that species. Less pronounced resonances at $\delta \approx 67$\,MHz and $\delta \approx 85$\,MHz likely correspond to the two Gallium isotopes, $^{69}$Ga and $^{71}$Ga. This confirms our ability to excite nuclear magnon modes in this Faraday configuration, as a first step towards a quantum memory~\cite{Appel2025}.

The presence of nuclear resonances is further confirmed by performing Hahn Echo spectroscopy~\cite{bechtold_three-stage_2015,Stockill2016}, where our probe sequence consists of the addition of a refocusing $\pi$ pulse half-way between the Ramsey pulses of Fig.\,\ref{fig4:ramsey}(a).
Fig.~\ref{fig5a:magnon}(b) shows the contrast as a function of total delay time $T$. The overall decaying contrast as a function of $T$ is fitted to a Gaussian function to extract the electron-spin qubit decoherence time $T_2^\text{HE} =1.14(2)$\,$\mu$s -- shorter than previous results on GaAs~\cite{Zaporski_Ideal_2023,shofer_tuning_2024}, which is consistent with a stronger interaction with the transverse nuclear field owing to a QD whose physical size is smaller (blue-shifted~\cite{Schimpf_Optical_2025}) and whose electron-spin splitting $\omega_e^0$ is smaller. Zooming in on short delays, shown in the inset of Fig.\,\ref{fig5a:magnon}(b), we see oscillations of the contrast, which a Fourier transform of the signal, shown in Fig.\,\ref{fig5a:magnon}(c), confirms to be predominantly at the As Larmor frequency. These Hahn Echo oscillations and the magnon resonances are features of an interaction between the electron spin and a precessing transverse nuclear field~\cite{Botzem2016,shofer_tuning_2024}, which allows the electro-nuclear energy exchange required for nuclear-spin cooling~\cite{jackson_optimal_2022} and quantum gates between the electron-spin qubit and a nuclear magnon~\cite{Appel2025}.

For an InGaAs QD, strain produces a large misalignment of nuclear and electronic quantization axes that, independently of the magnetic field direction due to strain inhomogeneity, naturally leads to a \emph{non-collinear} hyperfine interaction~\cite{Hogele2012} as used in cooling\cite{Gangloff_quantum_2019}. For a strain-free GaAs QD, a non-collinear interaction can arise from electronic g-factor anisotropy when the magnetic field is not aligned with the crystal axes of GaAs~\cite{shofer_tuning_2024}. When subject to a growth-axis magnetic field, however, as is the case here, the electronic and nuclear quantization axes should be fully aligned and there should not be a significant non-collinear interaction. Instead, the small electron-spin splitting $\omega_e^0 = 2.6$\,GHz makes the transverse \emph{collinear} interaction, whose amplitude scales as $\omega_e^{-1}$, sufficiently strong to be observed here. A hallmark of this interaction is that it enables only spin-conserving processes and thus only a single sideband in the ESR spectrum -- unlike the non-collinear interaction, which allows bidirectional sideband processes~\cite{Hogele2012, Gangloff_quantum_2019}. This dominant sideband is precisely what we observe in Fig.\,\ref{fig5a:magnon}b, where its blue-detuning ($\delta > 0$) is seen to be independent of the probe electron-spin state as consistent with a single spin-conserving process. We note the presence of a weaker red-detuned ($\delta < 0$) sideband near the As frequency (Fig.\,\ref{fig5a:magnon}a,b) -- suggesting a small non-collinear interaction is present, possibly due to residual strain or to a slight misalignment of the magnetic field with respect to the device -- which allows some degree of bidirectional feedback under the Raman cooling configuration (Fig.~\ref{fig4:ramsey}), despite the asymmetry~\cite{Gangloff2021}.

\begin{figure*}
    \centering
    \includegraphics[width=1\textwidth]{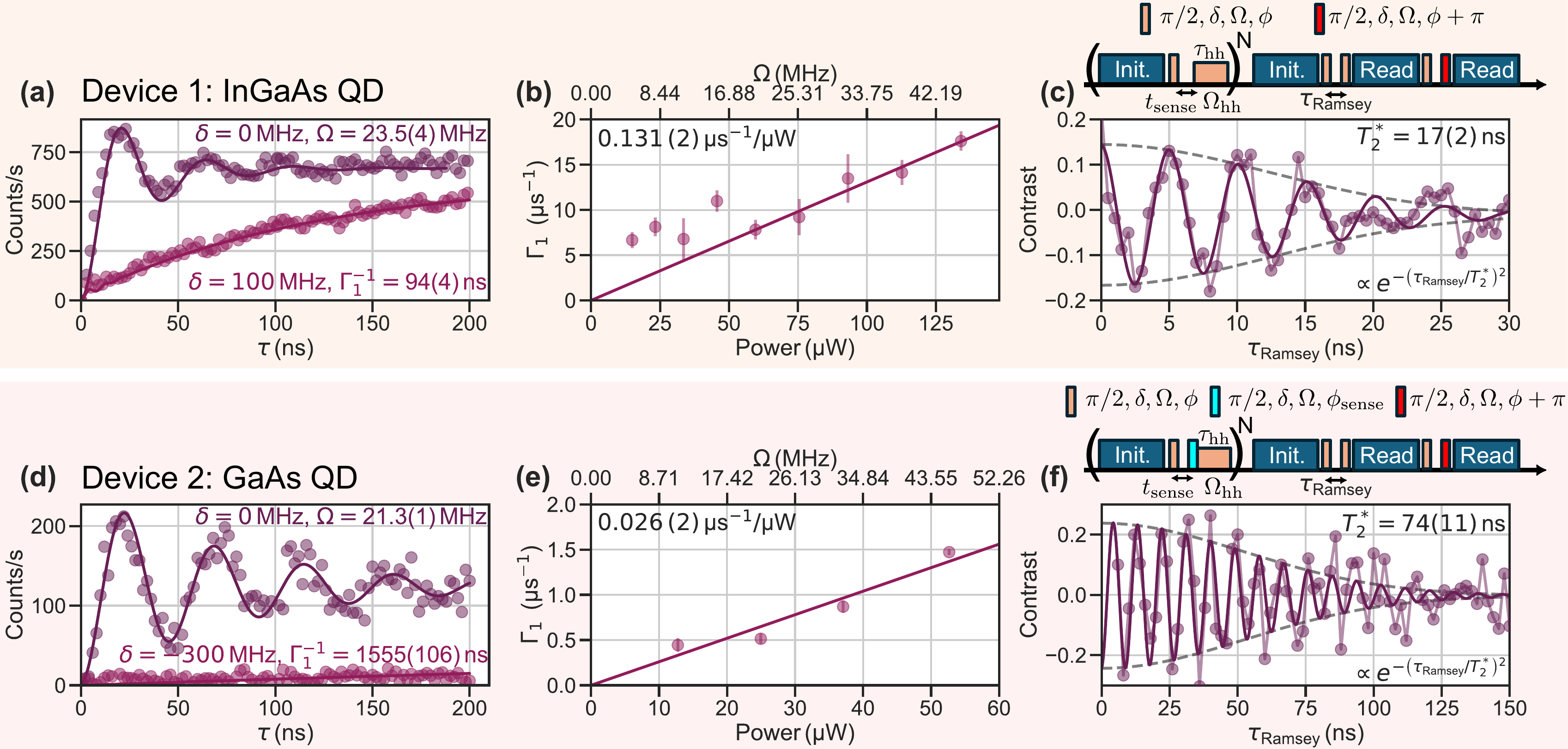}
    \caption{\textbf{Faraday quantum control of an electron-spin qubit in an InGaAs QD and a GaAs QD.} 
    \textbf{(a - c)} (\textbf{(d - f)}) are data for an InGaAs QD in Device 1 (a GaAs QD in Device 2). Details on quantum-algorithmic cooling used with Device 1 can be found in Appendix~\ref{app:ingaas_ramsey}. 
    \textbf{(a)} (\textbf{(d)}) Using the pulse sequence in Fig.\,\ref{fig3:rabi}a with quantum-algorithmic cooling (Raman cooling), Rabi oscillations at $\delta=0$ and laser-induced relaxation at $\delta=100$\,MHz ($\delta=-300$\,MHz). Solid curves are from our master equation model (Appendix \ref{app:model}) from which we fit $\Omega= 23.5(4)\,$MHz ($\Omega= 21.3(1)\,$MHz) and $\Gamma_1^{-1}=94(4)$\,ns ($\Gamma_1^{-1}=1555(106)$\,ns). \textbf{(b)} (\textbf{(e)}) $\Gamma_1$ as a function of Raman laser power. The solid line is a linear fit (no offset) with slope $0.131(2)\mu$s$^{-1}/\mu$W ($0.026(2)\mu$s$^{-1}/\mu$W). \textbf{(c)} (\textbf{(f)}) Our pulse sequence consists of quantum-algorithmic cooling, which for Device 2 (f) is adapted to the transverse collinear interaction by adding a $\pi/2$ pulse with a phase $\phi_\mathrm{sense}$ (see Appendix \ref{modified_ramsey_cooling}). This is followed by a Ramsey measurement as done in Fig.\,\ref{fig4:ramsey} but with a serrodyne frequency $f_\text{serr}=200$\,MHz (100~MHz) applied to the second Ramsey pulse. The data show Ramsey contrast as a function of Ramsey delay $\tau_\text{Ramsey}$. 
    The solid curve is a fitting function $A\sin(2\pi f\tau)\exp(-(\tau/T_2^*)^2)$ with free parameters: $A$ as the contrast amplitude, $f\approx f_\mathrm{serr}$ as the contrast oscillation frequency, and $T_2^*$ as the characteristic decay time. 
    The fit results are $T_2^* = 17(2)$\,ns ($T_2^* = 74(11)$\,ns),
    $f=(197(2)\,$MHz
    ($f=112(1)\,$MHz). Error bars indicate one standard error on fitted parameters.
    }
    \label{fig6:comparison}
\end{figure*}

\section{Control of QD Spins}

Lastly, we offer a succinct performance overview of InGaAs (Device 1) and GaAs (Device 2) QDs in Fig.~\ref{fig6:comparison}, where Rabi oscillations and laser-induced relaxation rate (at the same Rabi frequency), and Ramsey interferometry are shown side-by-side. Laser-induced relaxation -- which we know from previous work to be purely power-dependent and not related to scattering from the trion~\cite{Bodey_Opitcal_2019,Farfurnik2023} -- is measured as a function of Raman laser power and fitted to $0.131(2)$\,$\mu$s$^{-1}/\mu$W for Device 1 (which is $\sim 25\%$ of what was previously measured on the same wafer~\cite{Bodey_Opitcal_2019}) and to $0.026(2)$\,$\mu$s$^{-1}/\mu$W for Device 2, which is to the best of our knowledge the lowest value reported for any InGaAs or GaAs spin-control experiment. While the lower rate of laser-induced relaxation for Device 2 is a contributor to higher quality Rabi oscillations, the dominant dephasing mechanism for both devices is a $T_2$ process that likely arises in part from laser intensity noise coupled in via the differential Stark effect.

For Device 1, our best Rabi oscillations reach $\Omega=66(7)\,$MHz and $Q=0.90(8)$, and our best measured coherence is $T_2^* = 17(2)$\,ns with quantum-algorithmic cooling (further details in Appendix\,\ref{app:ingaas_ramsey}). For Device 2, our best Rabi oscillations reach $\Omega=273(6)\,$MHz and $Q=18.8(3)$, and our best measured coherence is $T_2^* = 74(11)$\,ns with quantum-algorithmic cooling -- in this case we adapted the cooling algorithm to use the transverse collinear interaction, as further described in Appendix~\ref{modified_ramsey_cooling}. 
While it is tempting to make conclusions about InGaAs vs GaAs QDs on this basis, it is unclear what leads to the striking difference in performance measured here -- it may simply be a difference in the optical quality of the QD heterostructure rather than a difference intrinsic to InGaAs vs GaAs QDs.

\section{Conclusions}

With this, we have demonstrated full quantum control of an electron-spin qubit in Faraday geometry for both InGaAs and GaAs QDs, where the cyclicity of spin readout transitions was measured to be $\mathcal{C}=291\,(28)$ and $\mathcal{C}=471\,(50)$, respectively. In this configuration, we have showed nuclear-spin cooling to enhance the electronic inhomogeneous dephasing time $T_2^*$ to tens of nanoseconds, which is sufficient to resolve and address nuclear magnons towards quantum-memory operation~\cite{Appel2025}. We note that while tens of nanoseconds of ground-state $T_2^*$ coherence is relatively short in comparison with other solid-state platforms~\cite{Debroux2021,Ruskuc2021a}, this enhanced value is two orders of magnitude longer than the radiative lifetime of a QD trion (hundreds of picoseconds even in the absence of Purcell enhancement). Combined with spin rotations which are not limited by nuclear-spin noise (Fig.\,\ref{fig3:rabi}b), our results are thus compatible with high-fidelity spin-photon entanglement~\cite{Meng2024}. As compared with the state-of-the-art in Voigt geometry of $T_2^*> 600\,$ns~\cite{nguyen_enhanced_2023} and $Q=45$~\cite{Appel2025} on an electron spin in a GaAs QD, we attribute the lower spin coherence and rotation fidelity to the coupling of the qubit splitting to laser intensity fluctuations via the differential Stark shift and to the increased relative importance of laser-induced relaxation in the presence of a weaker spin-flipping transition. Further improvements include stabilizing the laser intensity at the QD position to better than $0.1\,\%$ via vibration isolation and optimized power stabilization. Additionally, applying a vectorial magnetic field (i.e. a combination of Faraday and Voigt configurations) or a controlled few-percent strain to increase HH-LH mixing (on a GaAs QD~\cite{Chekhovich2018, Yuan2018}) would increase the strength of the spin-flipping transition and reduce the induced differential Stark shift, while keeping the cyclicity sufficiently high for single-shot readout. Single-shot measurements are not currently possible in our system due to a low end-to-end efficiency of less than $1\,\%$ after propagating through the microscope, optical filters, and single-photon detector. With a cyclicity of $\approx 500$, an improvement to $\sim10\%$ efficiency would suffice, made straightforward by photonic structures now reaching 70\% end-to-end efficiency with similar QDs~\cite{ding_-demand_2016,Uppu_Scalable_2020,tomm_abright_2021}, which combined with our method would allow for the generation of photonic cluster states with tens to hundreds of photons. 

\section*{Data Availability}
Data described in this paper and presented in the Supplementary Information are openly available~\cite{cam_repository}.

\section*{Author Contributions}
Z.X.K., U.H., B.D., A.M.H., D.G. and Y.K. carried out the experiments. 
E.C. and M.H. designed and fabricated QD Device 1. 
C.S., A.M.H., A.J.G., M.P., and A.R. designed the heterostructure for Device 2.  
A.J.G., M.P., and A.R. fabricated QD Device 2. 
J.M.K, M.G., and D.E.R developed the theoretical models and performed numerical simulations on the Rabi flops.
Z.X.K., J.M.K., and D.A.G. analyzed the data. 
M.A. and D.A.G. conceived and supervised the experiment.
Z.X.K. and D.A.G. wrote the manuscript with input from all the authors.

\section*{Competing Interests}
The authors declare no competing interests.

\begin{acknowledgments}
We acknowledge support from an EPSRC New Investigator Award EP/W035839/2 (D.A.G., Z.X.K.), the QuantERA project MEEDGARD through EPSRC EP/Z000556/1 (D.A.G., M.A.), ETRI through The Institute of Information \& communications Technology Planning \& Evaluation (IITP) grant funded by the Korea government (MSIT) (No.2022-0-00463, Development of a quantum repeater in optical fiber networks for quantum internet) (C.S., D.A.G.), the EU H2020 Research and Innovation Programme under the Marie Skłodowska-Curie grant QUDOT-TECH (861097; U.H., M.A.) and under Grant Agreement No. 871130 (Ascent+; A.R.), the EU HE EIC Pathfinder challenges action under grant agreement No. 101115575 (Q-ONE; A.R.), the Linz Institute of Technology Secure and Correct Systems Lab, supported by the State of Upper Austria (A.R.) and the FFG (Grants No. 891366 and 906046; A.R.), the Austrian Science Fund (FWF) by the research grant 10.55776/J4784 (C.S.) and via the Research Group FG5 (A.R.), the cluster of excellence quantA [10.55776/COE1; A.R.], MEEDGARD project (QuantERA II Program -- EU Horizon 2020 GA No.\ 101017733 and the National Centre for Research and Development, Poland project No.\ QUANTERAII/2/56/MEEDGARD/2024; M.G.), the German Federal Ministry of Research, Technology and Space
(BMFTR; grant number 16KIS2058), project MEEDGARD (J.M.K, D.E.R.). A.H. acknowledges an EPSRC DTP studentship through EP/W524311/1. B.D acknowledges an EPSRC DTP studentship through EP/D103586K/1. D.G. acknowledges the Undergraduate Research Opportunities (UROP) program at the University of Cambridge Cavendish Laboratory. D.A.G acknowledges a Royal Society University Research Fellowship. The InGaAs QD Device (Device 1) was grown in the EPSRC National Epitaxy Facility. 
\end{acknowledgments}

\appendix
\section{Experimental Setup}\label{app:hardware}
Device 1 (Device 2) is held at 4~K (3.4~K) inside a helium bath cryostat (closed-cycle cryostat) where a superconducting magnet produces a field $B_z$ up to 8~T in the $z$-direction. An objective lens with a Numerical Aperture of $0.5$ (Thorlabs C240TMD-B) collects light from Device 1, while an objective lens with a Numerical Aperture of $0.7$ (Thorlabs C330TMD-B) collects light from Device 2. A resonant laser and a Raman laser are combined on a beam splitter before they are sent into the cryostat.
The QD emission is then coupled back into an optical fiber before it is detected on an avalanche photodiode (APD) and time-tagged with a Swabian Time Tagger.
We employ cross-polarization in the photon collection path to reject the resonant readout laser, along with spectral filtering using a transmission grating setup (30\,GHz full-width-at-half-maximum) to further reject the Raman laser.

The resonant laser is linearly-polarized to enable equal excitation (and collection) of both the circularly-polarized dipoles of the trion $X^{1-}$.
We use an acousto-optical modulator (AOM) to pulse and actively stabilize the resonant laser power during experiments.

The Raman laser system consists of a pair of red-detuned lasers with equal intensities, derived from electro-optical modulation of a single narrow-linewidth laser.
This is done using a fiber-coupled amplitude-electro-optical modulator (EOM) which is locked to its interferometric minimum, and is driven by microwave pulses generated from an arbitrary waveform generator (AWG).
Due to a large $\omega_e$ ($>20$~GHz) in Device 1 (with InGaAs QDs), we use an IQ mixer (Marki Microwave, MMIQ-0218LXPC) to upconvert the output from two channels of an AWG (Tektronix 70002B) to $\simeq 15$~GHz with a microwave signal generator (Rohde and Schwarz, SMF100) as local oscillator.
In the case of Device 2 (with GaAs QDs), due to the smaller $\omega_e$ ($<5$~GHz), we directly synthesize the microwave pulses in the time domain using a single channel of an AWG (Real Digital, RFSoC 4x2, up to 9.85~GSPS) instead.
We build upon the \textit{qick} firmware~\cite{Stefanazzi_The_QICK_Quantum_2022} and develop a Python library (\textit{qeg-rfsoc}~\cite{danieljamesgraham_2025_17041953}) to interface with the board.
To minimize the photo-refractive effect in our EOM that causes large fluctuations in the extinction ratio and drift in the EOM bias, we seed the EOM with a laser power of $\sim10\,$mW.
We place an AOM after the EOM to pulse and actively stabilize the intensity of the Raman laser.
We then amplify the signal with an optical amplifier (Thorlabs, BOA780) and spectrally filter the amplified signal with a grating filter (full-width-at-half-maximum of 35~GHz) to remove the broad amplified spontaneous emission from the amplifier.
The filtered laser output is then coupled into an optical fiber before propagating through the Raman path. 
A Swabian Pulse Streamer is used to trigger the AWG and Swabian Time Tagger 20 and to generate optical pulses via electrical pulses to the AOM drivers.

\section{Device Properties and Spin $T_1$}\label{app:opticalspin}

\begin{figure*}
    \centering
    \includegraphics[width=1\textwidth]{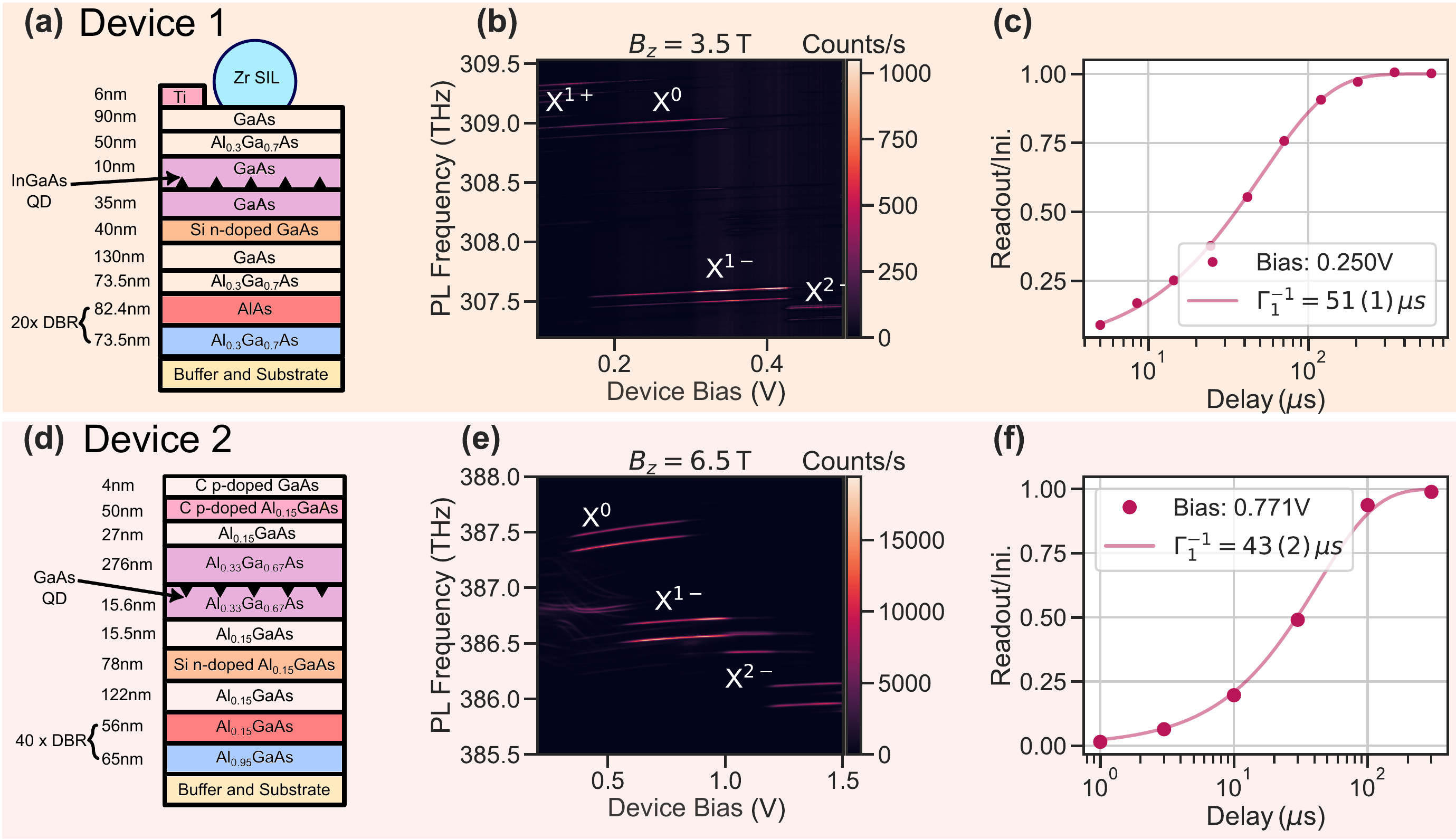}
    \caption{
    \textbf{(a)-(c)} (\textbf{(d)-(f)}) are measurements taken on Device 1 (Device 2). (a) and (d) are device heterostructures. (b) and (e) are photoluminescence emission frequency as a function of Device bias for above-band excitation. (c) and (f) are electron-spin relaxation experiments consisting of a resonant spin initialization (ini.) pulse, a variable delay, and a resonant spin readout pulse. The data points are the ratio of readout to initialization counts. The solid curves are fits to the data with an exponential function whose characteristic time is $\Gamma_1^{-1}$, as reported in the legend. Error bars indicate one standard error on fitted parameters.
    }
    \label{appendixa}
\end{figure*}

Device 1 consists of self-assembled InGaAs QDs embedded in a Schottky diode structure for deterministic charge control via the applied gate voltage.
The heterostructure is presented in Fig.\,\ref{appendixa}(a), taken from Refs.~\cite{stockill_phase-tuned_2017,Gangloff_quantum_2019, Jackson_quantum_2021}.
The gate voltage induces a d.c. Stark shift and tunes the emission wavelength of the emitter.
This is depicted in the photoluminescence data presented in Fig.~\ref{appendixa}(b) where we observe the presence and the splitting of the emission lines from the neutral ($X^0$) and charged excitons ($X^{1-}$, $X^{1+}$ and $X^{2-}$) at $B_z=3.5\,$T when exciting the QD with an above-band laser at 780~nm.
There is a 35-nm tunnel barrier between the n-doped layer and the QDs, and a capping layer above the QD layer to prevent charge leakage.
A planar cavity consisting of 20 pairs of distributed Bragg reflectors (DBR) and a super hemispherical cubic zirconia solid immersion lens (SIL), which effectively augments the Numerical Aperture of the objective from 0.5 to 1.15~\cite{Stockill}, are used to maximize photon out-coupling efficiency to $\sim 10\,\%$ for QDs with an emission wavelength around 970~nm.

Device 2 consists of Al-droplet-etched GaAs QDs, grown via molecular beam epitaxy, in a p-i-n diode. The heterostructure is presented in Fig.\,\ref{appendixa}(d).
To increase the photon out-coupling, the heterostructure contains 40 pairs of DBR mirror.
By applying a weak forward bias, we can deterministically charge the QD and induce a d.c. Stark shift to the emission wavelength.
We observe clear splitting of the emission lines at $B_z=6.5\,$T when excited with a 632~nm above-band laser. The photoluminescence emission as a function of device bias is shown in Fig.~\ref{appendixa}(e).

We estimate the electron-spin lifetime, $\Gamma_1^{-1}$ by performing time-resolved spin pumping experiments on the $X^{1-}$ transition.
By varying a delay between two resonant readout pulses, we can measure the readout counts normalized to the initialization counts as a function of delay.
We extract a $\Gamma_1^{-1}$ of $51(1)\,\mu s$ and $43(2)\,\mu s$ for Devices 1 and 2, respectively, by fitting an exponential function to the counts as a function of delay.
These results are shown in Fig.\,\ref{appendixa}(c) and \ref{appendixa}(f).

\begin{figure}
    \centering
\includegraphics[width=1\linewidth]{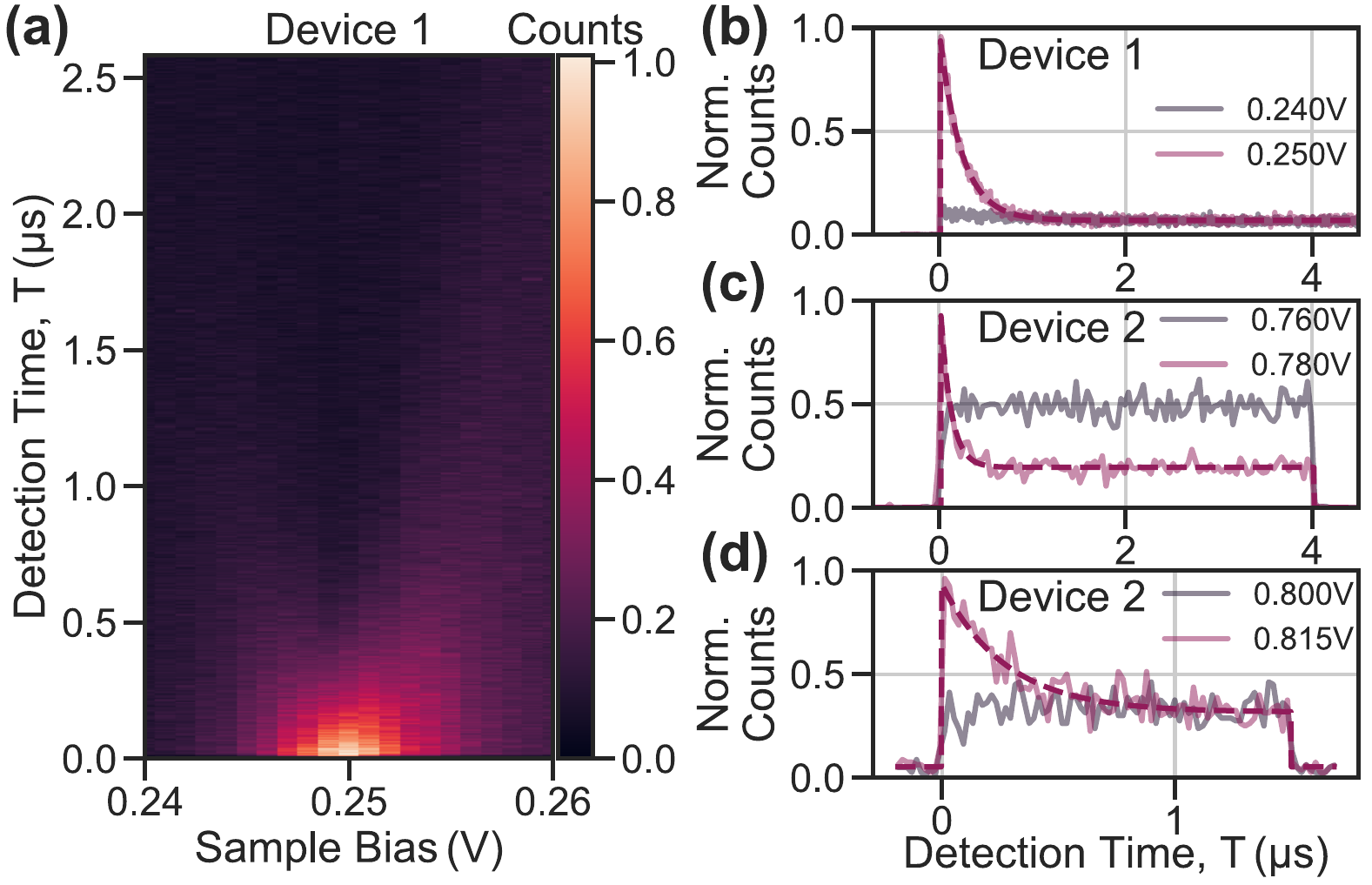}
    \caption{\textbf{(a)} Photon arrival histogram, from the $X^{1-}$ emission for the QD in Device 1 at $B_z=3.5\,$T, as a function of sample bias, for the initialization pulse. 
    \textbf{(b)} (\textbf{(c)}) Example histogram when the QD is bias-tuned to be resonant and off-resonant with the excitation laser in Device 1 (Device 2) with laser power of $\sim 16\,P_\mathrm{sat}$ ($\sim 6\,P_\mathrm{sat}$), measured at $B_z=3.5\,$T. The spin pumping time, obtained from the exponential fit on resonance at 0.250~V (0.780~V) is $\gamma_\mathrm{SP}^{-1}=234\,(2)\,$ns ($\gamma_\mathrm{SP}^{-1}=111\,(17)\,$ns).
    These data are from the same data set that generates the figures in Fig.~\ref{fig1:cyclicity}(d) and (e).
    \textbf{(d)} Example histogram when the QD is bias-tuned to be resonant and off-resonant with the excitation laser in Device 2 with a laser power of $\sim 0.5\,P_\mathrm{sat}$, measured at $B_z=6.5\,$T. The spin pumping time, obtained from the exponential fit on resonance at 0.815~V is $\gamma_\mathrm{SP}^{-1}=276\,(32)\,$ns, with the steady-state counts limited only by laser background. 
    % Spin initialization and readout on Device 1 and Device 2 are performed with a readout power similar to (b) and (d), respectively.
    }
    \label{fig:sp_ingaas_gaas}
\end{figure}

We estimate the spin-initialization fidelity from measuring the time-resolved photon arrival histogram during spin pumping, as a function of the QD-laser detuning, as set by the sample voltage bias via the DC Stark shift.
Fig.~\ref{fig:sp_ingaas_gaas}(a) shows an example of the photon arrival histogram as a function of sample bias for the QD in Device 1. 
We observe a shorter spin-pumping time when it is tuned to resonance. 
Examples of the photon arrival histograms at device biases corresponding to on and off-resonance on the QDs in Device 1 and Device 2, at powers well above saturation, are shown in Fig.~\ref{fig:sp_ingaas_gaas}(b) and (c), respectively.
For Device 1 (Fig.~\ref{fig:sp_ingaas_gaas}(b)), by subtracting the photon arrival histogram off-resonance from that on resonance, we can attribute the steady-state counts predominantly to the laser reflection background. For Device 2 (Fig.~\ref{fig:sp_ingaas_gaas}(c)), the steady-state counts can be attributed to repumping by the readout laser due to the small ground-state splitting and high optical power (with Rabi frequency comparable to the ground-state splitting). This is highlighted by the higher off-resonant steady-state counts (grey curve) compared to that on resonance in Fig.~\ref{fig:sp_ingaas_gaas}(c). For measurements on the GaAs QD in Device 2 presented in the main text in Figs.\,\ref{fig2:cooling}-\ref{fig6:comparison}, we thus use an optical power below saturation ($\sim 0.5P_{\rm sat}$) and a higher magnetic field (6.5~T) for spin initialization and readout to circumvent this error. This is shown in Figure~\ref{fig:sp_ingaas_gaas}(d), where the steady-state counts at a bias corresponding to being far off-resonance, where they are dominated by laser background, match the steady-state counts at a bias corresponding to resonance. 

For the settings that lead to a steady-state limited by laser background, the spin pumping rate $\gamma_\mathrm{SP}$ can be extracted from the fits (Fig.~\ref{fig:sp_ingaas_gaas}(b) and (d)), and one can calculate an upper bound to the spin-pumping fidelity~\cite{Kroner2008}, limited by $\Gamma_1$ and the spin pumping rate $\gamma_\mathrm{SP}$, of $F_\text{sp}^\text{upper}= (1+4\Gamma_1/\gamma_\mathrm{SP})^{-1}=98\%$ ($97\%$) for QDs in Device 1 (Device 2).

\section{Heavy-Hole Light-Hole Mixing and Raman Selection rules}\label{app:hhlh}

\subsection{Hole subband mixing}\label{app:hhlh-mixing}

We follow the HH-LH mixing treatment in the presence of strain from Ref.\,\cite{Wang2014}, which includes both direct first-order coupling between subbands and second-order terms involving the remote spin-orbit split-off band. In addition, we consider the spatial dependence of band admixtures. The relevant terms for understanding the spin-flipping transitions involve the valence subbands $\ket{J,J_z}$, where $J\in\{3/2,1/2\}$ and $J_z$ is the projection along the growth axis. We ignore any direct contributions of the split-off band $J=1/2$ to the eigenstates, which are negligible in comparison with mixing within the $J=3/2$ manifold, where $J_z=\pm1/2$ and $J_z=\pm3/2$ are the Light-Hole (LH) and Heavy-Hole (HH) valence subbands, respectively. However, we retain second-order virtual coupling via the split-off band, which yields additional mixing contributions in the $J=3/2$ manifold. 

Note that (solely) in this Appendix~\ref{app:hhlh-mixing}, we work in the \emph{valence electron} picture. When such a valence-electron state is removed, the corresponding \emph{hole} spinor used elsewhere in the manuscript is defined as its time-reversed spinor: $\ket{i}_{\mathrm{h}}=\mathcal{T}\ket{i}$, where $\mathcal{T}\ket{J,J_z}=(-1)^{J-J_z}\ket{J,-J_z}$.

In a nanostructure, electron states are generally spatially-dependent superpositions of bands,
\begin{equation}
	\ket{\psi(\vec{r})}=\sum_{\mu}\psi_{\mu}(\vec{r})\ket{\mu},
\end{equation}
where $\mu$ labels the bands [that also include the Conduction (C) band with $J = 1/2$ in addition to the mentioned valence bands] and $\psi_{\mu}$ are the distinct spatial envelopes for each band. We will further expand the envelopes in some basis of orbital states (e.g., 3D harmonic oscillator orbital states),
\begin{equation}
	\psi_{\mu}(\vec{r})=\sum_i a_i^{(\mu)} \phi_i^{(\mu)}(\vec{r}).
\end{equation}
For C-band electron states, we can safely neglect any band mixing and assume simple
\begin{equation}
	\ket{\uparrow\!\!/\!\!\downarrow}=\phi_0^{(\uparrow{/}\downarrow)}\Ket{\frac{1}{2},\pm\frac{1}{2}}
\end{equation}	
states modulated by spatially even ground-state envelopes. In contrast, band mixing in the valence band is generally strongly position-dependent (via different envelopes for different subbands), so it is impossible to represent a valence electron state as a position-independent ket.

Let $H_v$ be the $6x6$ valence-band Hamiltonian expressed in the $\ket{J,J_z}$ basis. Represented in our product basis of confined states, $\{\phi_{i}\ket{\mu}\}$, it takes the form
\begin{equation}
	H_v =
	\left(
	\begin{array}{ccc}
		\langle H_v \rangle_{00} & \langle H_v \rangle_{01} & \cdots \\
		\langle H_v \rangle_{10} & \langle H_v \rangle_{11} & \cdots \\
		\vdots                  & \vdots                  & \ddots
	\end{array}
	\right),
\end{equation}
where $\langle O \rangle_{ij} = \int \dd^3\vec{r}\, \phi_i^*(\vec{r}) \, O \,\phi_j(\vec{r})$ and thus $\langle H_v \rangle_{ij}$ are $6\times6$ blocks for each combination of the envelope basis functions, describing the coupling between the envelope sectors $i$ and $j$. Then, we look for LH contributions to the predominantly HH hole ground state that are generated by the off-diagonal elements of this Hamiltonian between $\phi_0(\vec{r})\ket{3/2,\pm3/2}$ and some $\phi_i(\vec{r})\ket{3/2,\pm1/2}$ for any $i$. Thus, a given total LH admixture has an envelope $\psi_{\mu}$ composed of many envelope basis functions $\phi_i$ of different symmetry.

For that reason, it is impossible to provide a simple representation of the hole states, as band mixing is generally position-dependent. However, for optical transitions, we are only interested in the transition dipole moment matrix elements taken between conduction and valence (hole) states, $\braket{c|\vec{d}|v}$, where $c=\uparrow,\downarrow$ and $v$ stands for the valence (time-reversed hole) eigenstate. Under our assumption on the conduction states, this becomes
\begin{align}
	\vec{d}_{cv} \simeq{}& \sum_{\mu} \int \dd^3\vec{r} \, \phi_0^*(\vec{r}) \, \psi_{\mu}(\vec{r}) \braket{c|\vec{d}|\mu} \notag\\
	={}& \bra{c} \vec{d} \, \sum_{\mu} \int \dd^3\vec{r} \, \phi_0^*(\vec{r}) \, \psi_{\mu}(\vec{r})\ket{\mu}, \notag\\
	\equiv{}& \braket{c|\vec{d}|V},
\end{align}
which shows that for optical transitions only the projections of the hole mixing envelopes onto the even electron envelope matter.
This allows us to write such projected, position-independent `effective' lowest valence states associated with the lowest hole doublet, $\ket{V}= \int \dd^3\vec{r} \, \phi_0^*(\vec{r}) \, \ket{v}$ as:
\begin{align}\label{eqn:holestates}
    \Ket{\Uparrow} &= \alpha\Ket{\frac{3}{2},+\frac{3}{2}} + \chi\Ket{\frac{3}{2},+\frac{1}{2}} + \epsilon\Ket{\frac{3}{2},-\frac{1}{2}},\notag\\
    \Ket{\Downarrow} &= \alpha\Ket{\frac{3}{2},-\frac{3}{2}} - \chi^*\Ket{\frac{3}{2},-\frac{1}{2}} + \epsilon^*\Ket{\frac{3}{2},+\frac{1}{2}},
\end{align}
where $\alpha \lesssim 1$ parametrizes the dominant HH contribution (without loss of generality, we take $\alpha$ to be real and thus our phase reference), and $\chi$ and $\epsilon$ to be small complex numbers parameterizing the optically-active part of the LH contribution. To relate this to the hole picture, we use the convention that the hole spinor associated with a missing valence electron is the time-reversed spinor of that electron. Thus, e.g., a missing $\Ket{\Uparrow}$ electron corresponds to $\Ket{\Downarrow}_{\mathrm{h}}=\mathcal{T}\Ket{\Uparrow}$, which has formally the same spinor expansion as $\Ket{\Downarrow}$. In spin-preserving transitions, the contribution of the $\chi$ term is dark, while $\epsilon$ creates a small opposite circular polarization contribution leading to ellipticity of the transition dipoles. The spin flipping transitions $\Uparrow_{\mathrm{h}} \rightarrow\downarrow$ and $\Downarrow_{\mathrm{h}}\rightarrow\uparrow$ allowed by the $\chi$ term are circularly polarized and by the $\epsilon$ term $\pi_z$ polarized.

For $H_v$, we combine the Luttinger-Kohn valence-band Hamiltonian with the Bir-Pikus Hamiltonian accounting for strain. We then get 
\begin{align}
	\chi = \int \dd^3\vec{r} \, \phi_0^*(\vec{r}) \, \chi(\vec{r}), \notag \\
	\epsilon = \int \dd^3\vec{r} \, \phi_0^*(\vec{r}) \, \epsilon(\vec{r}),
\end{align}
with
\begin{align}\label{app:mixing}
	\chi(\vec{r}) =& \sum_i \frac{-\sqrt{2}\Delta_\mathrm{SO}\langle S\rangle_{0i}^* - \sqrt{6}\langle S\rangle_{0i} \langle R \rangle_{0i}^*}{\langle \Delta_{\mathrm{lh}} \rangle_{0i} \Delta_\mathrm{SO}} \, \phi_i(\vec{r}),\notag\\
	\epsilon(\vec{r}) =& \sum_i \frac{(\frac{3}{2}\langle \Delta_{\mathrm{lh}} \rangle_{0i} + \Delta_\mathrm{SO})\langle R \rangle_{0i}^* + \sqrt{3}(\langle S\rangle_{0i}^*)^2}{\langle \Delta_{\mathrm{lh}} \rangle_{0i} \Delta_\mathrm{SO}} \, \phi_i(\vec{r}),
\end{align}
where
\begin{align}\label{app:birpikus_full}
%	\Delta_{lh}(\vec{r}) =& -\frac{\hbar^2}{2m_0}\,2\gamma_2\!\left(k_z^2 - \frac{k_x^2 + k_y^2}{2}\right) \notag\\
%	&+ \frac{b}{2}\left[e_{xx}(\vec{r})+e_{yy}(\vec{r})-2e_{zz}(\vec{r})\right],\notag\\
	R(\vec{r}) =& -\frac{\hbar^2}{2m_0}\sqrt{3}\!\left[\gamma_2 (k_x^2 - k_y^2)
	- 2i\gamma_3 k_x k_y\right] \notag \\ 
	&+ \frac{\sqrt{3}}{2}b\left[e_{xx}(\vec{r})-e_{yy}(\vec{r})\right] - i d e_{xy}(\vec{r}),\notag\\
	S(\vec{r}) =& -\frac{\hbar^2}{2m_0}2\sqrt{3}\,\gamma_3 (k_x - i k_y) k_z \notag\\
	&+ \frac{d}{\sqrt{2}}\left[e_{zx}(\vec{r})-i e_{yz}(\vec{r})\right]
\end{align}
are the relevant off-diagonal HH-LH mixing matrix elements of $H_v$ in the band basis, and
\begin{align}\label{app:birpikus_full2}
	\Delta_{\mathrm{lh}}(\vec{r}) ={}& E_{\mathrm{HH}}^{(0)}-E_{\mathrm{LH}}^{(0)} = -\frac{\hbar^2}{2m_0}\,2\gamma_2\!\left(k_z^2 - \frac{k_x^2 + k_y^2}{2}\right) \notag\\
	&+ \frac{b}{2}\left[e_{xx}(\vec{r})+e_{yy}(\vec{r})-2e_{zz}(\vec{r})\right],\notag\\
\end{align}
is the splitting between the HH and LH band edges ($E_{\mathrm{HH}}^{(0)}$ and $E_{\mathrm{LH}}^{(0)}$) in the absence of band mixing. Here, 
$\Delta_\mathrm{SO}$ is the spin-orbit coupling constant, 
$e_{ij}$, $i,j\in\{x,y,z\}$, is the strain tensor,  
$m_0$ is the free-electron mass, 
$k_i$ ($i\in\{x,y,z\}$) are the components of the crystal-momentum operator, 
and $\gamma_2$, $\gamma_3$ are the Luttinger parameters characterizing the valence-band effective masses.

Let us first analyze the structure of terms in Eqs.~\eqref{app:mixing} and \eqref{app:birpikus_full}, focusing on differences between GaAs and InAs QDs and temporarily neglecting the position dependence. Hole subband splitting, $\Delta_{\mathrm{lh}}$, has contributions due to confinement and biaxial strain. For larger and almost unstrained GaAs QDs, both contributions are typically small compared to spin-orbit interaction, hence, in terms of magnitude, $\Delta_{\mathrm{lh}} \ll \Delta_\mathrm{SO}$, $|SR^*| \ll |\Delta_\mathrm{SO}S^*|$ and $|S^2|\ll |R|$. As a result, we can initially expect $\chi \approx -\sqrt{2}S^*/\Delta_{\mathrm{lh}}$ and $\epsilon \approx R^*/\Delta_{\mathrm{lh}}$. For InGaAs QDs, however, the confinement is stronger, and also no strain contribution can be ignored, as strain energy scales can be comparable to spin-orbit coupling; as a result, no such simplification can generally be made to these terms.

On top of that, symmetry of the envelopes comes into play via parity selection rules for the electron and hole envelopes (momentum conservation relaxed by confinement). From Eq.~\eqref{app:birpikus_full}, we learn that both $R$ and $S$ have kinetic and strain contributions. The kinetic part of $R$ is even and therefore can only produce even contributions to LH admixture envelopes. Additionally, it requires in-plane anisotropy of the QD confinement potential. Thus, it is typically present in InAs QDs, where shear strain and piezoelectric field break the rotational symmetry of confinement even in otherwise symmetric QDs, leading to significant ellipticity of allowed transitions and significant $\pi_z$ polarized contribution to forbidden transitions. Apart from the QD shape, engineered external fields that deform the hole wave function could be used to tune this effect. On the other hand, in less strained and close to rotationally symmetric GaAs QDs, this contribution can be negligible. The strain part of $R$ relies on the in-plane shear strain, which is a sign-changing field around the dot and predominantly couples the ground state to the admixture envelopes of odd in-plane symmetry. As a result, GaAs QDs with low biaxial strain and resultant low HH-LH subband splitting can still have a significant $\epsilon$ admixture, which is, however, odd and thus does not contribute to the allowed optical transitions. In InAs QDs, this strain contribution to $R$ is present but does not affect the allowed transitions.

Inspecting $S$, we can see that its kinetic part is odd along the growth axis, since it involves $k_z$. Any admixture this part could produce would therefore require an envelope excited along that direction, which strongly suppresses this contribution in both types of dots, especially in flat InAs QDs. The important part thus comes from the shear strains in the vertical planes. These, again, are odd-like functions and thus couple to odd envelopes. However, $S$ enters the equation for $\epsilon$ squared and in the one for $\chi$ multiplied by $R$ in one of the terms. The term $SR$ is therefore most likely responsible for the spatially even and bright contribution to the $\chi$ admixture in both types of QDs. It also produces a minor contribution to the bright part of $\epsilon$. Thus, shear-strain engineering seems to be the route to tune this admixture, while biaxial strain compensating for the built-in HH-LH splitting can enhance both types of admixtures, especially in GaAs QDs.

Up to this point, we neglected the impact of the magnetic field on HH-LH subband mixing. The difference in Faraday-geometry $g$-factors for HH and LH valence bands, $g_{\mathrm{HH}} = 6\kappa+27 q/2$, $g_{\mathrm{LH}} = 2\kappa+q/2$ (with $\kappa=1.28$, $q=0.04$ in GaAs \cite{Mlinar2005}), leads to different values of $\Delta_{\mathrm{lh}}$ when different spin configurations are considered, with the Zeeman contribution being
\begin{equation}\label{eq:app-HHLHsplitting}
\Delta_{\mathrm{lh;\Uparrow/\Downarrow,\uparrow/\downarrow}}^{(\mathrm{Z})} = \frac12 \left( \pm g_{\mathrm{LH}} \mp g_{\mathrm{HH}} \right) \mu_{\mathrm{B}}B_z,
\end{equation}
and the $\pm$ symbols referring independently to the orientations of the LH- and HH-subband spins. Note that signs in Eq.~\eqref{eq:app-HHLHsplitting} need to be flipped when hole states are considered. In InGaAs QDs, this can be neglected in view of large splitting due to strain, but in GaAs QDs, these Zeeman contributions can be significant and lead to unequal LH admixtures to the states $\ket{\Uparrow}$ and $\ket{\Downarrow}$ in Eq.~\eqref{eqn:holestates}. In particular, unequal values of $\chi$ lead to the difference in the imbalances $\eta$ on the two $\Lambda$ subsystems.

\subsection{Effect on Raman Transitions}

For a general Raman laser polarization (where, as in our experiment, both Raman fields are the same polarization), we have two lambda systems, corresponding to two trion intermediate states, operating in parallel. The first in which the trion state with $\Downarrow$ links a spin-conserving transition in absorption, $(\Omega_\downarrow^\mathrm{sc})^*$, and a spin-flipping transition in emission, $\Omega_\uparrow^\mathrm{sp}$. The second in which the trion state with $\Uparrow$ links a spin-flipping transition in absorption, $(\Omega_\downarrow^\mathrm{sp})^*$, and a spin-conserving transition in emission, $\Omega_\uparrow^\mathrm{sc}$. The total Raman coupling between the two ground states is then:

\begin{equation}
    \Omega = \frac{(\Omega_\downarrow^\mathrm{sc})^*\Omega_\uparrow^\mathrm{sp}}{2\Delta} + \frac{(\Omega_\downarrow^\mathrm{sp})^*\Omega_\uparrow^\mathrm{sc}}{2(\Delta+\omega_h)}
\end{equation}

Let us first consider Raman fields with a polarization in the $(x,y)$ plane, for which $|\Omega_{\sigma_-}|^2$ and $|\Omega_{\sigma_+}|^2$ are proportional to laser intensity in the $\sigma_-$ and $\sigma_+$ modes, respectively. In this situation, only the $\chi$ term LH mixing (Eq.\,\ref{eqn:holestates}) is relevant, as it is the one that allows in-plane polarized spin-flipping transitions. The coupling between the ground states becomes:
\begin{equation}
    \Omega = \frac{-\chi_{\Downarrow}^*|\Omega_{\sigma_-}|^2}{2\Delta} + \frac{\chi_{\Uparrow}^*|\Omega_{\sigma_+}|^2}{2(\Delta+\omega_h)},
\end{equation}
where we made explicit the possible difference of $\chi$ in the magnetic field. We see immediately that for $|\Omega_{\sigma_-}|^2 = |\Omega_{\sigma_+}|^2$, i.e. linearly polarized light, and in the limit of $B_z\to 0$ ($\omega_h \to 0, \chi_{\Downarrow}=\chi_{\Uparrow})$, the two lambda systems destructively interfere and $\Omega \rightarrow 0$.
Including the field, we still deal with $\chi_{\Downarrow}\approx \chi_{\Uparrow}$ for InGaAs QDs, and the significant $\omega_h$ makes the coupling nonvanishing. In contrast, we already deduced $\chi_{\Uparrow}>\chi_{\Downarrow}$ for GaAs QDs, which compensates for $\omega_h$ in the denominator, keeping $\Omega$ close to vanishing, as we see for a GaAs QD (Fig.\,\ref{fig2:cooling} and Appendix \ref{app:polarization}). 

If we introduce a coupling of the Raman fields to $\pi_z$ -- which is possible when focusing our laser through an objective with large numerical aperture taking us away from the paraxial limit -- and parametrize it as a fraction $\eta$ of a linearly polarized field $\Omega_{\pi}$ in-plane, then the coupling between the ground states becomes:
\begin{equation}
    \Omega = \frac{\Omega_{\sigma_-}^*(-\chi^*\Omega_{\sigma_-}+\epsilon^*\eta\Omega_{\pi})}{2\Delta} + \frac{\Omega_{\sigma_+}(\chi\Omega_{\sigma_+}^* +\epsilon^*\eta\Omega_{\pi}^*)}{2(\Delta+\omega_h)}.
\end{equation}
Here, a deviation from pure circular polarization would activate the $\pi_z$ contribution and could enhance the Raman coupling. This would require a significant value of the $\epsilon$ LH mixing term arising, for example, from in-plane asymmetry of the confinement, and in-plane shear strains. InGaAs QDs exhibit both, and this mechanism could thus explain why the optimal Raman coupling for these QDs is elliptical, as seen in Appendix \ref{app:ingaas_ramsey}. The solid immersion lens present in Device 1 further increases the numerical aperture and could enhance this effect.

\section{Device 2 Raman Polarization Maps Supplementary Data}\label{app:polarization}
\begin{figure*}
    \centering
    \includegraphics[width=1\textwidth]{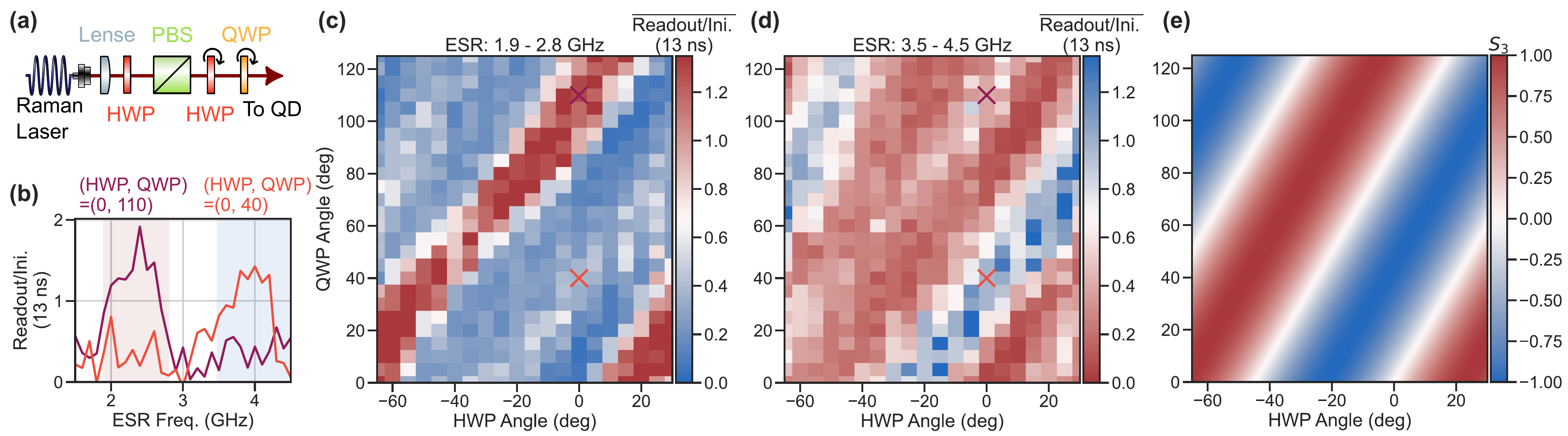}
    \caption{\textbf{Raman laser polarization control.} All measurements are taken on a GaAs QD in Device 2.
    \textbf{(a)} Schematic of the optical path for the Raman laser, highlighting the components used for polarization control. PBS: polarizing beam splitter, HWP: half-wave plate, and QWP: quarter-wave plate.
    \textbf{(b)} ESR spectrum at (HWP, QWP) = (0, $40^o$) and (HWP, QWP) = (0, $110^o$) showing a shift in the ESR resonance from the bare electron-spin splitting at $B_z=6.5~T$ of $\omega_e^0\approx 2.6\,$GHz.
    \textbf{(c)} \textbf{(d)} Mean readout counts (normalized to the initialization counts) as a function of HWP and QWP angles in the ESR frequency range between 1.9 and 2.8~GHz (3.5 and 4.5~GHz). 
    \textbf{(e)} Degree of circular polarization, quantified by the $S_3$ Stokes parameter as a function of HWP and QWP angles.
    $S_3=1(-1)$ yields $\sigma_{-}(\sigma_{+})$ polarization.
    }
    \label{appendixb}
\end{figure*}

The Raman fields used in our experiment consist of a pair of red-detuned lasers with equal intensities, derived from electro-optical modulation of a single narrow-linewidth laser.
Fig.~\ref{appendixb}(a) shows the experimental setup of the Raman laser path where we control the polarization of the Raman laser via a pair of motorized half-wave (HWP) and quarter-wave (QWP) plates.
To calibrate the angles of the HWP and the QWP, we place a polarizing beam splitter (PBS) after the two wave plates and by iteratively changing the angle of both waveplates to minimize the laser light at the reflection port of the PBS, we identify the angles that correspond to the linear horizontal polarization (i.e. (HWP, QWP) = (0, 0)).

To find the optimal optical polarization, we take the ESR spectrum, by scanning the microwave frequency applied to the electro-optical modulator and detecting the readout counts, as a function of the polarization of the Raman laser pair.
We employ the following pulse sequence: nuclear-spin cooling (Rabi cooling), spin initialization with a resonant pulse, probing with a Raman drive of a fixed pulse duration of $13\,$ns, and spin readout with a resonant pulse.
We keep the laser power fixed and the cooling and probe frequencies equal to each other, $\omega_\text{cool} = \omega$, for all of the scans.
We first fix the angle of the HWP to $0^o$ while iteratively varying the angle of the QWP.
We observe a variation in the readout counts, along with a shift in the resonance as we vary the QWP angle.
Fig.~\ref{appendixb}(b) shows the two extreme cases, where we observe a maximal negative shift and a maximal positive shift from the bare electron splitting of $\omega_e^0\approx 2.6\,$GHz at (HWP, QWP)=($0^o$, $40^o$) and (HWP, QWP)=($0^o$, $110^o$), respectively. 
These polarization settings correspond to the dipoles of the two spin-conserving transitions, $\sigma_{-}$ and $\sigma_{+}$, respectively, which are also identified from polarization-resolved photoluminescence measurements. 

Scanning both the HWP and QWP angles while recording the readout counts confirms the dependence on the polarization of the shift in the ESR resonance.
The bright regions in Fig.\,\ref{appendixb}(c) and Fig.\,\ref{appendixb}(d) denote the mean readout signal within the two frequency ranges where we found the ESR resonances in Fig.\,\ref{appendixb}(c) (1.9-2.8~GHz and 3.5-4.5~GHz, respectively).
We find a striking resemblance in the results depicted in Fig.\,\ref{appendixb}(c) with the variation in the degree of circular polarization, quantified by the Stokes parameter $S_3$, as a function of HWP and QWP angles, as shown in Fig.\,\ref{appendixb}(e).
We observe a maximum red-shifted (blue-shifted) signal in the region where $S_3=1$ ($S_3=-1$), corresponding to the $\sigma_-$ ( $\sigma_+$) polarization.
These results confirm that, for Device 2, the optimal polarization for the Raman fields is the polarization aligned to the dipole of the $X^{1-}$ transition.

\section{Intensity Noise Model and $Q$}\label{app:intensitynoise}
\begin{figure*}
    \centering
    \includegraphics[width=1\textwidth]{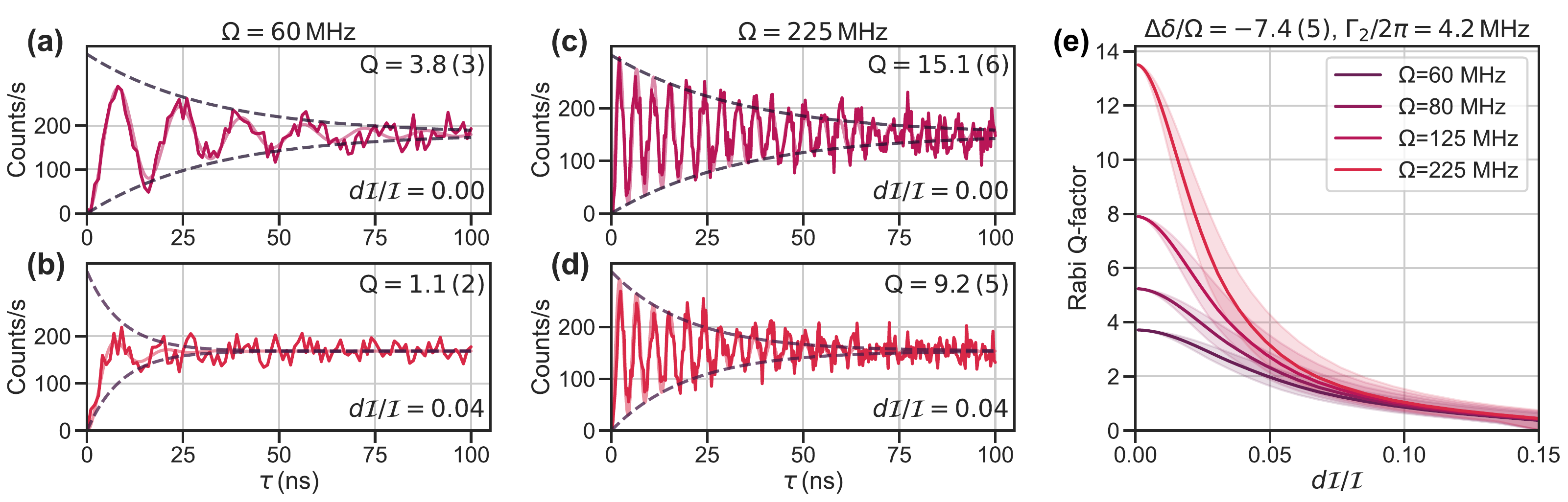}
    \caption{\textbf{Influence of laser intensity noise on the quality factor $Q$ of Rabi oscillations.} All measurements are taken on a GaAs QD in Device 2.
    \textbf{(a, b)} Same measurement sequence as in Fig.\,\ref{fig3:rabi}. Readout counts as a function of pulse duration $\tau$ following Raman cooling ( $\Omega_\mathrm{cool}\approx 15\,$MHz) at fractional intensity fluctuation, $d\mathcal{I}/\mathcal{I}$ of $
    0.00$ (\textbf{(a)}) and $0.04$ (\textbf{(b)}) at the same probe Rabi frequency of $\Omega\approx 60\,$MHz. Solid curves are fits to the data using a two-level master equation model from which we obtain $Q$. 
    \textbf{(c, d)} Same as (a, b) but at $\Omega\approx 225\,$MHz.
    \textbf{(e)} Simulated variation in $Q$ as a function of $d\mathcal{I}/\mathcal{I}$ at various probe Rabi frequencies $\Omega$. 
    Here, $\Delta\delta/\Omega=7.4(5)$ and $\Gamma_2/2\pi=4.2\,$MHz. The shaded regions indicate three standard error on $\Delta\delta/\Omega$.
    }.
    \label{appendixc}
\end{figure*}

To quantify the effect of laser intensity noise on the Rabi oscillation quality factor, we systematically introduce intensity modulation on the Raman laser by modulating the voltage applied to the acousto-optical modulator (AOM) used for laser intensity stabilization.
We do so by combining the output of a white noise source (derived from an arbitrary waveform generator Rigol DG832) with the output of our PID controller with a power splitter before sending them to the analog modulation port of the AOM driver.
This is to ensure we have a constant noise power spectral density (up to the bandwidth of the 35~MHz waveform generator) and can explore the effect of noise from low to high frequency simultaneously.
A sample-and-hold protocol is implemented at the PID controller during the experiment to ensure minimal feedback from the PID controller on the external noise.
By calibrating the response of the AOM on the laser power, we can convert the peak-to-peak modulation voltage to fractional (rms) intensity fluctuation $d\mathcal{I}/\mathcal{I}$. Fig.~\ref{appendixc}(a,c) and \ref{appendixc}(b,d) show the readout counts as a function of pulse duration $\tau$ at laser noise values of $d\mathcal{I}/\mathcal{I}$ of $0.00$ and $0.04$, respectively. 
Fig.~\ref{appendixc}(a, b) (c, d) are taken at $\Omega=60$\,MHz ($\Omega=225$\,MHz).
The measurements are done with the same cooling Rabi frequency, $\Omega_\mathrm{cool}\approx 15\,$MHz as in Fig.\,\ref{fig3:rabi}. The added intensity noise clearly degrades the quality of Rabi oscillations.

To understand and fit these dynamics, we use a two-level model calculated in \textit{Qutip}~\cite{qutip1, qutip2}.
We run a master equation solver, which includes collapse operators describing the decoherence mechanisms of spin relaxation $\Gamma_1$ (fixed with values taken from the result in Fig.\,\ref{fig6:comparison}(e), i.e. $\Gamma_1=\alpha\,\Omega, \alpha=0.0048(3)$) and spin dephasing $\Gamma_2$ (free parameter). We also include the effect of slow detuning noise ($T_2^*$ fixed to the Ramsey measurement value of Fig.\,\ref{fig4:ramsey}(b)) by repeating the master equation simulation for a set of detunings, and performing an average of the output of the simulation weighted with a normal distribution whose standard deviation $\sigma$ is defined by both the laser intensity fluctuation $d\mathcal{I}/\mathcal{I}$ and the nuclear Overhauser noise $\sigma_{OH}$ i.e.
\begin{align}
    \sigma_{laser} &= d[\Delta \delta] = (\Delta\delta/\Omega) \, d\Omega =(\Delta\delta/\Omega) \,\Omega \,d\mathcal{I}/\mathcal{I},\\
    \sigma_{OH}&=\sqrt{2}/T_2^*,\\
    \sigma &= \sqrt{\sigma^2_{laser} + \sigma^2_{OH}}.
\end{align}
To simplify the fits over a range of $d\mathcal{I}/\mathcal{I}$ values, we first fit the measurement data (Rabi flops) at $d\mathcal{I}/\mathcal{I}=0$ for each dataset to extract $\Omega$ (in the absence of any external laser-induced detuning noise).
We find an average spin dephasing rate of $\Gamma_2/2\pi = 4.2(4)\,$MHz for all 3 measurement settings. We fix $\Gamma_2$ to this value, and for each value of $\Omega$, we then run the simulations as a function of $d\mathcal{I}/\mathcal{I}$. The ratio of differential Stark shift to the probe Rabi frequency is fixed to the mean experimental value $\Delta \delta/\Omega =7.4$ found in Fig.\,\ref{fig2:cooling}(f).

Solid curves in Fig.\,\ref{appendixc}(a-d) are fits to our two-level model described above, allowing us to extract the Rabi quality factor from the $\pi$-pulse contrast $f_\pi$ via $Q=-1/\ln(2f_\pi-1)$, obtained from the fitted density matrix,
showing a clear decrease in $Q$ as $d\mathcal{I}/\mathcal{I}$ increases. By varying the magnitude of $d\mathcal{I}/\mathcal{I}$ in our simulation, we obtain the full set of theoretical predictions, shown as solid curves in Fig.\,\ref{appendixc}(d), for four values of $\Omega$. 
These are consistent with the experimental data (also shown in Fig.\,\ref{fig3:rabi}), and thus indicate that the drop in $Q$ at increasing $d\mathcal{I}/\mathcal{I}$ can be explained by the coupling of intensity noise to the differential Stark effect.

\section{Modified Quantum Algorithmic Cooling}\label{modified_ramsey_cooling}
Figures~\ref{fig5a:magnon}(a,b) and \ref{fig6:comparison}(f) show the modified quantum-algorithmic cooling sequence we use to stabilize the ESR, from which we achieve an electronic $T_2^*=74(11)\,$ns.
In place of the original quantum-algorithmic cooling scheme~\cite{jackson_optimal_2022} (used with the InGaAs QD in Device 1), we add an extra $\pi/2$ pulse, whose phase is $\pi/2$ relative to the first $\pi/2$ pulse, following the sensing period of the algorithm and directly before the electro-nuclear interaction (hh) pulse. This additional pulse turns an electronic coherence along $\pm x$ to $\pm z$ to enable a nuclear spin flip conditioned on the electronic state with the single-sideband process allowed by the collinear hyperfine interaction (see Fig.\,\ref{fig5a:magnon}(b)).
The cooling algorithm uses the following parameters to obtain the data of Fig.\,\ref{fig5a:magnon}(a,b) and Fig.\,\ref{fig6:comparison}(f): fixed sensing time of $t_\mathrm{sense}=55\,$ns, $N=30$, $\Omega_{hh}\approx 47\,$MHz, $\tau_{hh}=60\,$ns, and $\phi_\mathrm{sense}=0.875\pi$. 
Note that $\phi_\mathrm{sense}$ is not $\pi/2$ here only owing to an uncompensated free precession phase. The $\pi/2-$pulses are done at $\Omega=125\,$MHz, corresponding to a duration of 2~ns.
The cooling sequence is dominated by the $1.5\,\mu s$-long resonant pulse to reset the spin, resulting in a total cooling duration of $\sim 55\,\mu s$. 

\section{InGaAs Supplementary Data}\label{app:ingaas_ramsey}
\begin{figure*}
    \centering
    \includegraphics[width=1\linewidth]{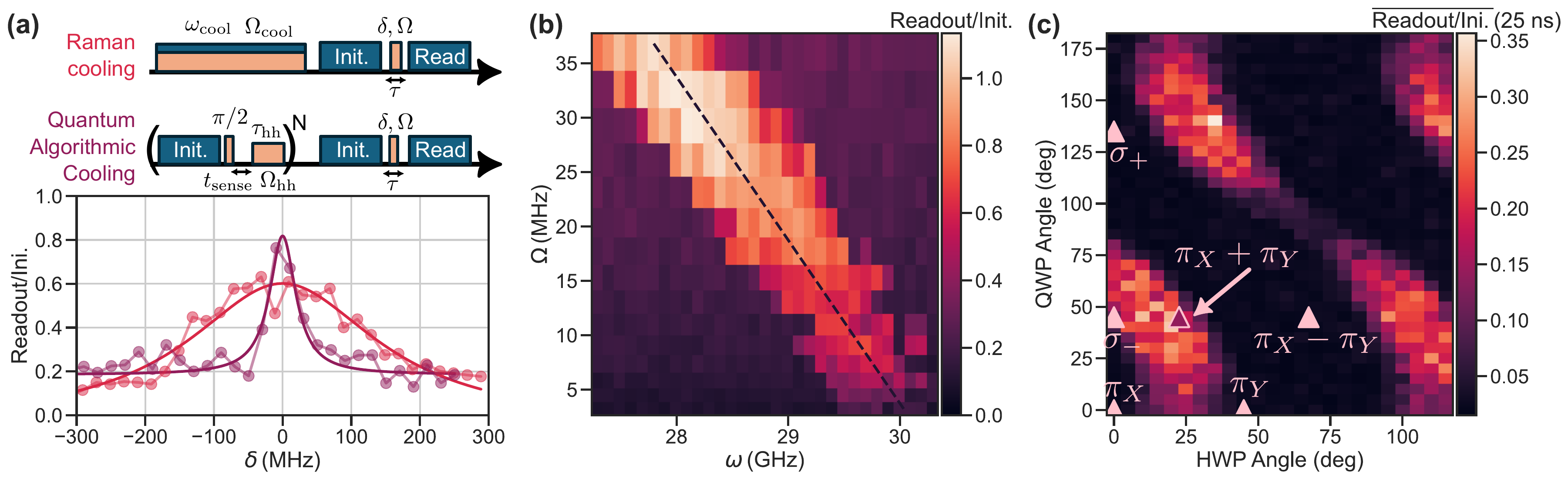}
    \caption{\textbf{Raman control and nuclear-spin cooling on an InGaAs QD in Faraday configuration.} All measurements are taken on an InGaAs QD in Device 1.
    \textbf{(a)}  ESR spectroscopy. Pulse sequences (top panel) show a cooling period, with either $50\,\mu s$ of Raman or $85\,\mu s$ of quantum-algorithmic cooling, followed by a probe period consisting of spin initialization ($2.5\,\mu s$), ESR drive, and spin readout ($2.5\,\mu s$). The duration ratio of probe to cooling is $\approx 10\,\%$. 
The bottom panel shows the resulting data with $\Omega\sim 30\,$MHz. Raman and quantum-algorithmic cooling are shown as red and purple data, respectively. A Gaussian fit (solid curves) yields linewidth (full-width-at-half-maximum) of $261(23)$\,MHz and $53(6)$\,MHz, respectively. At detuning $\delta$ well beyond the ESR feature, the constant value of the readout highlights the effect of spontaneous spin-flips due to optically-induced relaxation processes~\cite{Bodey_Opitcal_2019}.
    \textbf{(b)} Probe readout under Rabi cooling as a function of ESR drive frequency and Rabi frequency. The dashed line highlights a linear negative shift in the ESR resonance from the bare electron splitting of $\omega_e^0\approx30\,$GHz.
    \textbf{(c)} Maximum readout counts in an ESR spectrum at a fixed Raman laser power (corresponding to $\Omega \sim 20\,$MHz), under Rabi cooling, as a function of the polarization of the Raman laser, as controlled by rotating half-wave-plate (HWP) and quarter-wave-plate (QWP) angles (referenced to the linearly horizontal polarization $\pi_X$). Polarizations of interest are highlighted with triangles.
}
    \label{fig:spin_control_ingaas}
\end{figure*}

Fig.~\ref{fig:spin_control_ingaas}(a) shows the narrowing of the ESR resonance with Raman (top panel) and quantum-algorithmic cooling (middle panel) at $\Omega_\mathrm{cool}=\Omega\approx30\,$MHz, with a probe pulse duration set to the $\pi$-pulse time of $\tau=17\,$ns. 
A $2.5\,\mu s$ long resonant pulse is used for initialization and the readout pulses. 
For quantum-algorithmic cooling, we optimize the cooling parameters to be $\tau_{hh}=105\,$ns, $t_\mathrm{sense}$ chirped from 10~ns to 200~ns in $N=30$ steps, amounting to a duration of $\sim 85\,\mu s$. We achieve an ESR linewidth (at full-width-at-half-maximum) of $261(23)$\,MHz and $53(6)$\,MHz for Raman and quantum-algorithmic cooling, respectively.

We also observe the effect of the differential Stark shift on the ESR resonance, shown in Fig.\,\ref{fig:spin_control_ingaas}(b). We take ESR spectra as a function of $\Omega$, and observe a negative shift in the ESR resonance as a function of $\Omega$.
These data yield $\Delta\delta/\Omega=-65(2)$ from a linear fit, and a corresponding Rabi frequency imbalance of $\eta=129(5)$. 
We can then vary the polarization of the Raman laser pair to optimize the readout signal integrated over the ESR frequency range 25-35~GHz (i.e. $\pm 5\,$GHz around the bare ESR resonance of $\omega_e^0=30\,$GHz), as shown in Fig.\,\ref{fig:spin_control_ingaas}(c).
In contrast to our measurement on a GaAs QD (Fig.\,\ref{fig2:cooling}) where we found maximum readout signal with pure circular polarization, we find here that a degree of diagonal polarization $\pi_X + \pi_Y$ results in the highest readout signal.
This polarization setting is used for the electron-spin control measurements shown in Fig.\,\ref{fig:spin_control_ingaas}(a-b) and Fig.\,\ref{fig6:comparison}(a-c).

\section{Trion Lifetime} \label{app:lifetime}
Due to hardware limitations at the time of the experiments on Devices 1 and 2, we did not measure the trion lifetime of the QDs in these devices.
However, since Device 1 is the same as the one used in Refs.~\cite{Stanley_dynamics_2014, stockill_phase-tuned_2017}, we estimate the lifetime from statistics acquired in these references to be $696\,(65)\,$ps, based on the measured values of the lifetime, $584\,(10)\,$ps, $692\,(5)\,$ps, $735\,(10)\,$ps, $727\,(10)\,$ps and $742\,(10)\,$ps with QDs emitting at $\sim 970~$nm.

\begin{figure}
    \centering
    \includegraphics[width=1\linewidth]{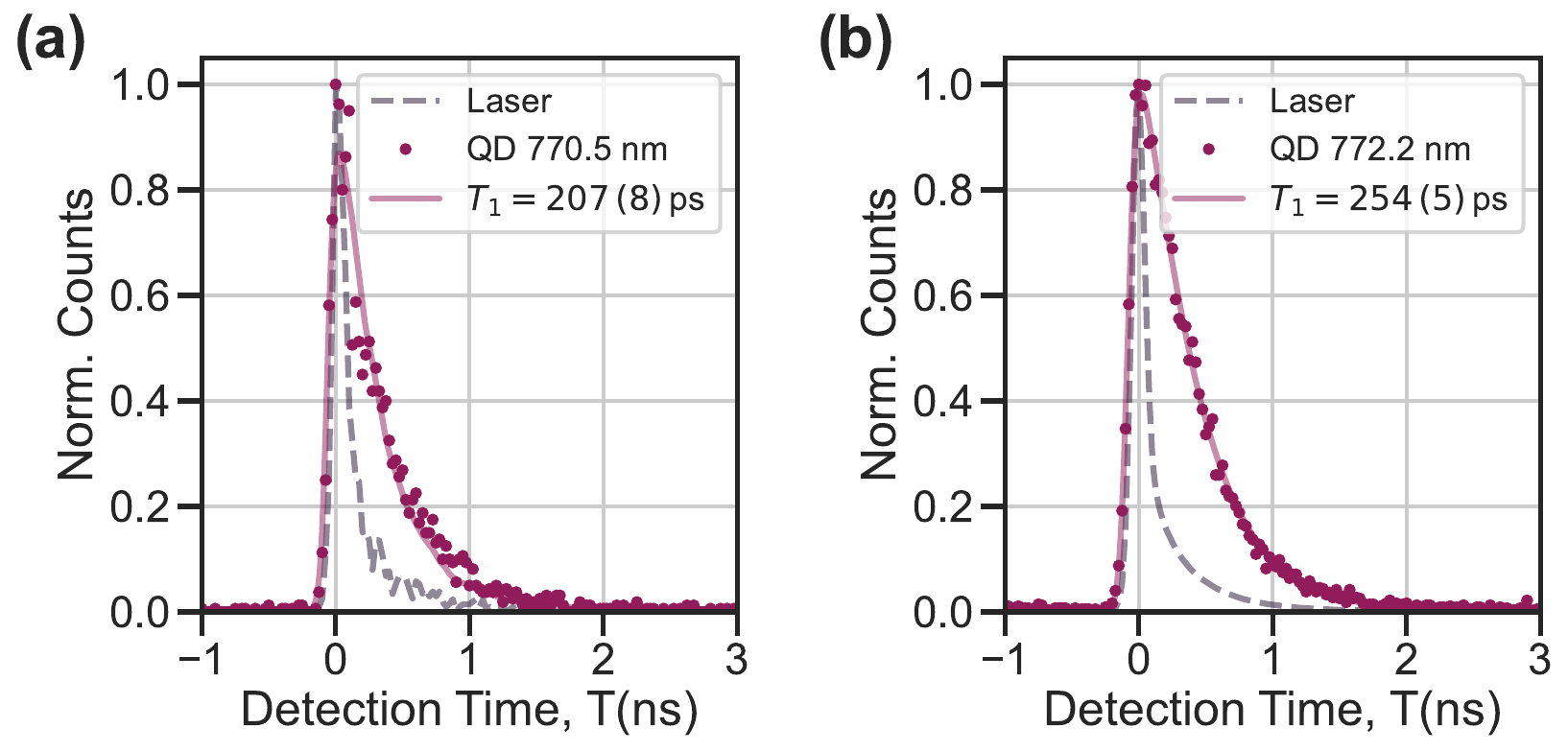}
    \caption{\textbf{Trion lifetime on Device 2.} \textbf{(a, b)} Trion lifetime, measured using a picosecond pulse laser, exciting resonantly and near resonant (phonon-assisted) in (a) and (b), respectively. The fits (solid-lines) are obtained by re-convoluting an exponential decay with the instrument-response function (dashed lines), giving a lifetime of $T_1=207\,(8)\,$ps and  $T_1=254\,(5)\,$ps for the data in (a) and (b) respectively.}
    \label{fig:lifetime_gaas}
\end{figure}

For Device 2, we estimate the trion lifetime of the QD used in the manuscript from measuring the trion lifetime of QDs from the same wafer (JKU Linz, SA1093).
Using a pico-second laser (Coherent MIRA 900-HP) as an excitation source and a fast detector (MPD, PD-020-CTE-FC) as the single-photon detector, we measure the trion lifetime from the QDs, one via resonance fluorescence, and two via longitudinal acoustic phonon-assisted excitation.
Ref.~\cite{Reindle_highly_2019} shows that longitudinal phonon-assisted excitation results in a similar lifetime compared to resonance fluorescence, allowing us to spectrally-separate single-photons from the stronger, broad pulse laser.
We measure the photon arrival histogram, synchronized to the trigger from the pulse laser, and fit an exponential decay, convoluted with the instrument response function, which we use to extract the lifetime.
We obtain a lifetime of 207~(8)~ps, 254(5)~ps, and 245~(8)~ps for QDs emitting at 770.5~nm, 772.2~nm, and 775.5~nm, respectively.
The latter two QDs are measured via longitudinal phonon-assisted excitation.
An example of the lifetime data is depicted in the Fig.~\ref{fig:lifetime_gaas}, highlighting the fits (solid curve), the instrument-response-function (dashed curve) and the experimental data (circle dots).
We obtain an estimate of 235~(25)~ps for the trion lifetime, from these three measured values.

\section{Three-level Model for CPT} \label{app:cpt}

We start with the Hamiltonian of a three-level system, 
\begin{align*}
    \mathcal{H}_{3LS}&=\hbar \Delta \ket{e}\bra{e} + \hbar \delta \ket{\uparrow}\bra{\uparrow} \\ &+ \hbar\Omega_{\downarrow}/2 \left(\ket{\downarrow}\bra{e} + \ket{e}\bra{\downarrow}\right) \\
    &+ \hbar\Omega_{\uparrow}/2 \left(\ket{\uparrow}\bra{e} + \ket{e}\bra{\uparrow}\right),
\end{align*}
where $\ket{\downarrow}$ and $\ket{\uparrow}$ are the two electron-spin ground states linked by a trion excited state $\ket{e} \equiv \ket{\Downarrow\downarrow\uparrow}$, $\delta=\omega-\omega_e^0$ is the detuning from two-photon resonance, and $\Delta$ is the single-photon detuning.
The resonant Rabi frequencies for each transition from the excited state are labeled as $\Omega_{\downarrow}$ or $\Omega_{\uparrow}$.
We include the collapse operators that define the ground-state relaxation $\Gamma_{1}$ and dephasing rate $\Gamma_2$, along with the spontaneous relaxation from the excited state along the spin-conserving $\gamma_\mathrm{SC} \approx \gamma_1$ and spin-flipping $\gamma_\mathrm{SP}$ channels  as $\hat{C}_1=\sqrt{\Gamma_{1}/2} \left(\ket{\downarrow}\bra{\uparrow} + \ket{\uparrow}\bra{\downarrow} \right)$,  $\hat{C}_2=\sqrt{\Gamma_2/2}\left(\ket{\downarrow}\bra{\downarrow}-\ket{\uparrow}\bra{\uparrow}\right)$, $\hat{C}_3=\sqrt{\gamma_{1}} \ket{\downarrow}\bra{e} $ and $\hat{C}_4=\sqrt{\gamma_\mathrm{SP}} \ket{\uparrow}\bra{e} $, respectively. 

To fit the data, we fix $\Delta=0$  and the ground and excited state relaxation rates to be $\Gamma_{1}^{-1}=45000\,$ns, $\gamma_{1}^{-1}=0.25\,$ns and $\gamma_\mathrm{SP}^{-1}=100\,$ns.
We then keep $\omega_e^0$, $\Omega_{\uparrow}$, $\Omega_{\downarrow}$ and $\Gamma_{2}$ as free fitting parameters, and $\omega$ as our control axis.
By fitting to the data in Fig.\,\ref{fig2:cooling}(b), we obtain $\omega^0_e/2\pi=2.60(1)\,$GHz, $\Omega_{\downarrow}=9.3(3)\,\mathrm{ns}^{-1}$, $\Omega_{\uparrow}=0.19(2)\,\mathrm{ns}^{-1}$ and $\Gamma_2=0.53(8)\,\mathrm{ns}^{-1}$. 
Given that the saturation Rabi frequency $\Omega_\mathrm{sat}=\gamma_{1}/\sqrt{2}=2.82\,\mathrm{ns^{-1}}$, we can represent both Rabi frequencies in terms of $\Omega_\mathrm{sat}$, such that $\Omega_{\downarrow}=3.29(9)\,\Omega_\mathrm{sat}$ and $\Omega_{\uparrow}=0.069(8)\,\Omega_\mathrm{sat}$.
The former is consistent with the pump power used in the experiment ($\approx 10\,\mathrm{P_{sat}}$, hence $\Omega_{\downarrow} \approx \sqrt{10}\,\Omega_\mathrm{sat}$) .

\section{Four-level Model for Raman-driven Rabi Oscillations}\label{app:model}

\begin{figure}[h]
    \centering
    \includegraphics[width=0.8\linewidth]{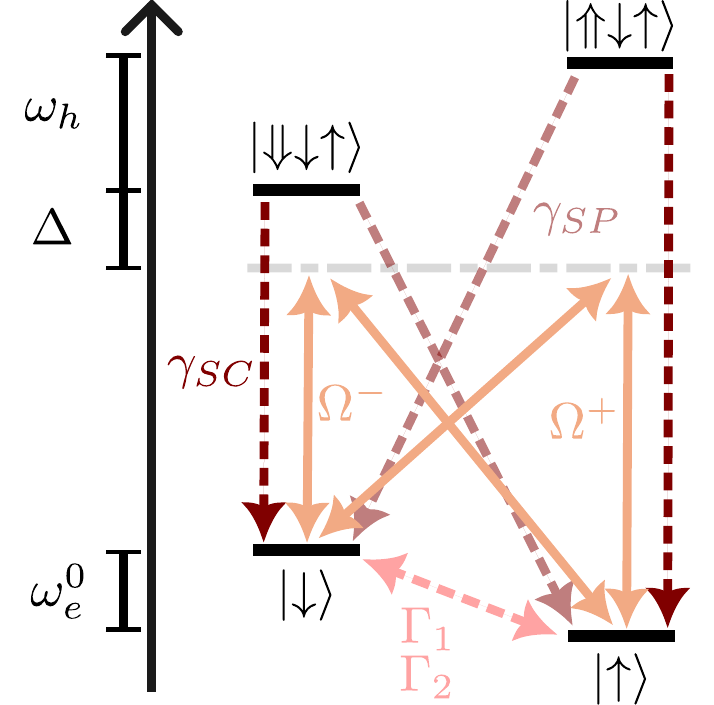}
    \caption{Four-level system.
    Solid arrows represent detuned Raman drive.
    % Pink arrows indicate optically active transitions. 
    Dashed arrows represent Lindblad operators, where we include the trion relaxation rates ($\gamma_{SC}$ and $\gamma_{SP}$), along with the ground state electron relaxation ($\Gamma_1$) and dephasing ($\Gamma_2$) rates.
    }
    \label{fig:level-scheme}
\end{figure}
We consider a four-level system consisting of two spin states $\ket{\uparrow}$ and $\ket{\downarrow}$ in the ground state manifold, where a single electron occupies the QD. In the excited state manifold we account for two $X^{1-}$ states $\ket{\Uparrow\downarrow\uparrow}$ and $\ket{\Downarrow\downarrow\uparrow}$. A sketch of the system is shown in Fig.~\ref{fig:level-scheme}.

In Faraday geometry, the Hamiltonian of the system can be separated into the system Hamiltonian $H_0$ and the part accounting for the light-matter interaction $H_\text{int}$ via
\begin{equation}
    H = H_0 + H_{\text{int}}\,.
\end{equation}
We set the energies of the states, such that $E_{\ket{\downarrow}}=0$, $E_{\ket{\uparrow}}=-\hbar \omega_e^0$ and $E_{\ket{\Uparrow\downarrow\uparrow}}=E_{\text{trion}}$ and $E_{\ket{\Downarrow\downarrow\uparrow}}=E_{\text{trion}}+\hbar\omega_h$, where the magnetic field has induced a Zeeman splitting of electrons ($\omega_e^0$) and holes ($\omega_h$). 
We write the Hamiltonian in a rotating frame defined by the frequency of the first laser pulse, $E_\text{trion}/\hbar - \Delta$, resulting in the detuning $\Delta$. In this frame,  the second laser pulse appears with an additional detuning $\Delta_\text{RF}$.
The bare Hamiltonian is then given by
\begin{align}
    \begin{split}
        H_0 &= 0\ket{\downarrow}\bra{\downarrow} -\hbar \omega_e\ket{\uparrow}\bra{\uparrow} \\ &+ \hbar\Delta \ket{\Downarrow\downarrow\uparrow}\bra{\Downarrow\downarrow\uparrow} 
        + \hbar\left(\Delta + \omega_h\right) \ket{\Uparrow\downarrow\uparrow}\bra{\Uparrow\downarrow\uparrow}.   
    \end{split}
\end{align}
The light-matter interaction in standard rotating wave and dipole approximation reads
\begin{align}
    \begin{split}
        H_{\text{int}} &= -\frac{\hbar}{2} \left(\Omega^{-} \ket{\Downarrow\downarrow\uparrow}\bra{\downarrow}
                          -\Omega^{+} \ket{\Uparrow\downarrow\uparrow}\bra{\uparrow}\right)   \\
                        & -\frac{\hbar}{2} \frac{1}{\sqrt{\mathcal{C}}} \left(\Omega^{-} \ket{\Downarrow\downarrow\uparrow}\bra{\uparrow}
                          + \Omega^{+} \ket{\Uparrow\downarrow\uparrow}\bra{\downarrow}\right) + h.c.
    \end{split}
\end{align}
The cyclicity $\mathcal{C}$ is used to tune the ratio between strong spin-conserving and weak spin-flipping transitions.
% We assume that we drive both transitions with a pair of linearly-polarized lasers, such that we achieve equal Rabi frequencies for $\mathcal{C}=1$.
%We choose to work in a rotating frame set to the ${\ket{\downarrow}} \rightarrow {\ket{\Uparrow\uparrow\downarrow}}$ transition frequency $E_\mathrm{trion}/\hbar - \Delta$. 
% In the chosen rotating frame, the second laser pulse appears with an additional detuning $\Delta_{\mathrm{RF}}$. 
The left ($\Omega^{-}$) and right ($\Omega^{+}$) circularly polarized components of the envelope are given by
\begin{equation}
    \Omega^{-} = \Omega_{1}+\Omega_{2}\exp^{-i\Delta_{RF}t}
\end{equation}
\begin{equation}
    \Omega^{+} = 0
\end{equation}
to match the experimental configuration. 

To describe relaxation processes from the trions to the ground states, we use Lindblad rates $\Gamma^\prime_i \in \{\gamma_{SC}, \gamma_{SP}\}$ and their corresponding Lindblad operators $L_i$. We also employ Lindblad terms to account for a direct relaxation between both ground states at rate $\Gamma_1$. Furthermore, we take into account the dephasing of electron ground state coherences by exponentially damping them with a constant damping rate $\Gamma_2$. These processes are shown in Fig.~\ref{fig:level-scheme} by dashed arrows.

This leads to the equations of motions for a density matrix 
% \begin{equation}
%     \dot{\rho} = -\frac{i}{\hbar}\left[H,\rho\right] + \sum_{i} \Gamma^\prime_i \left(L_i\rho L_i^\dagger - \frac{1}{2} \left\{ L_i^\dagger L_i,\rho\right\} \right) +\sum_{i\neq j}\gamma\rho_{ij}.
% \end{equation}
\begin{equation}
    \dot{\rho} = -\frac{i}{\hbar}\left[H,\rho\right] + \sum_{i} \Gamma^\prime_i \left(L_i\rho L_i^\dagger - \frac{1}{2} \left\{ L_i^\dagger L_i,\rho\right\} \right).
\end{equation}
We solve this equation numerically to calculate the dynamics of the system. 
% The parameters used for the calculations are given in Tab.~\ref{tab:sim-params}. 
% In addition, we consider that the QDs in an ensemble vary in geometry, and therefore the transition energies and g-factors differ slightly. To account for these variations, we determine the system's dynamics for different Zeeman splittings under the constraint $\omega_{h}=2\omega_{e}$. 
To include the effect of the finite dephasing, evident from the decay time scale ($T_2^*$ value) and the order (n=2) from the free-induction decay experiments, we perform a weighted average on $\rho(t, \omega)$, i.e.  
\begin{equation}
    \rho_{\text{ensemble}}(t) = \int w(\omega)\rho(t,\omega) \mathrm{d} \omega
\end{equation}
with the weight function modeled as a Gaussian noise with $\sigma=\sqrt{2}/T_2^*$, 
\begin{equation}
    w(\omega) = \frac{1}{\sigma\sqrt{2\pi}}\exp^{-\frac{1}{2}\left( \frac{\omega-\omega_{e}}{\sigma} \right)^{2}}
\end{equation}
is performed. 

\begin{figure*}
    \centering
    \includegraphics[width=1\linewidth]{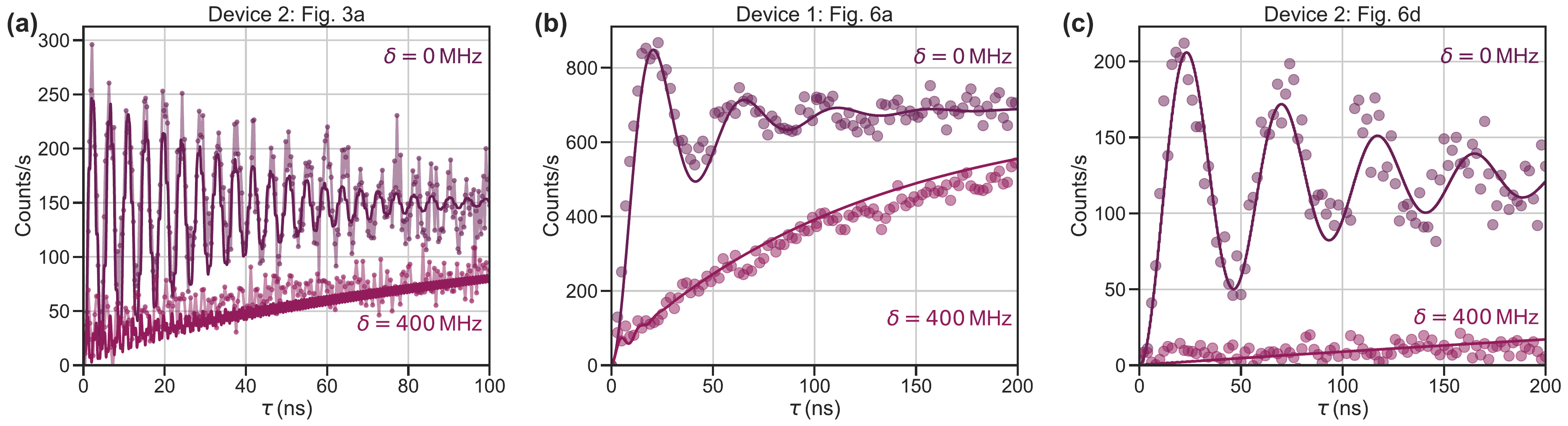}
    \caption{\textbf{Four-level Model Simulation.} \textbf{(a, b, c)} The output of the four-level model, on ($\delta=0$) and off-resonance ($\delta\neq0$), using the parameters in Table~\ref{table:4ls_fits} for the experimental data in Fig.~\ref{fig3:rabi}(a), \ref{fig6:comparison}(a), and \ref{fig6:comparison}(d), respectively.}
    \label{fig:4ls_simulation}
\end{figure*}

Using the parameters listed in Table~\ref{table:4ls_fits}, we can match the output of the simulation with the experimental data in Fig.\,\ref{fig3:rabi}(a) and \ref{fig6:comparison}(a, d).
As solving the four-level model is resource extensive, fitting of the experimental data in Fig.\,\ref{fig3:rabi} and \ref{fig6:comparison} is done with a two-level model~\cite{Bodey_Opitcal_2019, Gangloff_quantum_2019} instead.
Both models result in similar values of the Rabi frequency $\Omega$, ground state relaxation $\Gamma_1$ and pure dephasing $\Gamma_2$.

\begin{table*}[]
\label{table:4ls_fits}
\caption{Parameters used to produce the fits in Fig.\,\ref{fig3:rabi}(a), \ref{fig6:comparison}(a) and \ref{fig6:comparison}(d).}
\begin{tabular}{|l|l|l|l|l|l|l|l|}
\hline
\textbf{Dataset} & \textbf{$\Omega/2\pi$\,(MHz)} & \textbf{$\gamma_{SC}/2\pi$\,(MHz)} & \textbf{$\gamma_{SF}/2\pi$\,(MHz)} & \textbf{$\Gamma_1/2\pi$\,(MHz)} & \textbf{$\Gamma_2/2\pi$\,(MHz)} & \textbf{$T_2^*$\,(ns)} & \textbf{$\Delta/2\pi\,$(GHz)}\\ \hline
Fig.~\ref{fig3:rabi}(a)          &    227                                 &         637                           &       2.83                             &   0.106                                &        4.55                           &                       34   & 600    \\ \hline
Fig.~\ref{fig6:comparison}(a)         &        23                             &                       227             &         0.784                           &                      1.06          &                 5.68                     &                 17      & 600         \\ \hline
Fig.~\ref{fig6:comparison}(d)           &    21                                 &             637                       &         2.83                           &            0.106                       &          2.65                         &        34           & 600             \\ \hline
\end{tabular}
\end{table*}

\bibliography{reference}

\end{document}